\documentclass[twocolumn]{aastex631}

\usepackage{appendix}
\usepackage{graphicx}
\usepackage{subfigure}
\usepackage{lipsum}
\usepackage[utf8]{inputenc}
\usepackage{textcomp}
\usepackage{newunicodechar}
\newunicodechar{′}{\ensuremath{\prime}}
\let\longtable*\relax

\begin{document}

\title{A search for Be stars in multiple systems within the solar neighborhood }

\author{V. M. Kalari}
\affiliation{Gemini Observatory/NSF’s NOIRLab, Casilla 603, La Serena, Chile}
\author{R. Salinas}
\affiliation{Nicolaus Copernicus Astronomical Center, Polish Academy of Sciences, Bartycka 18, 00-716 Warszawa, Poland}
\author{C. S\'aez-Carvajal}
\affiliation{Instituto de F\'isica y Astronom\'ia, Universidad de Valpara\'iso,Av. Gran Bretaña 1111, 5030 Casilla, Valpara\'iso, Chile}
\affiliation{Gemini Observatory/NSF’s NOIRLab, Casilla 603, La Serena, Chile}
\author{R. D. Oudmaijer}
\affiliation{Royal Observatory of Belgium, Ringlaan 3, B-1180 Brussels, Belgium}
\affiliation{School of Physics \& Astronomy, University of Leeds, Woodhouse Lane, Leeds LS2\,9JT, UK}
\author{S. Howell}
\affiliation{NASA Ames Research Center, Moffett Field, CA 94035, USA}
\author{S. Caballero-Nieves}
\affiliation{Embry–Riddle Aeronautical University, Daytona Beach FL}
\author{K. Kamp}
\affiliation{Embry–Riddle Aeronautical University, Daytona Beach FL}
\author{R. Matson}
\affiliation{U.S. Naval Observatory, 3450 Massachusetts Avenue NW, Washington, D.C. 20392, USA}
\author{N. Scott}
\affiliation{The CHARA Array of Georgia State University, Mount Wilson Observatory, Mount Wilson, CA 91023, USA}
\author{T. Cao}
\affiliation{Department of Astronomy, Xiamen University, 422 Siming South Road, Xiamen 361005, People′s Republic of China}
\author{Z. Hartman}
\affiliation{NASA Ames Research Center, Moffett Field, CA 94035, USA}
\author{H. Kim}
\affiliation{Gemini Observatory/NSF’s NOIRLab, Casilla 603, La Serena, Chile}




\begin{abstract}
Be stars are widely considered to be the product of binary interaction. However, whether all Be stars are formed via binary interaction is unclear, and detailed estimates of the multiplicity of Be stars and characterization of their components are required. In this study, we present speckle observations of 76 Be stars taken using the Gemini North and South speckle imagers spanning angular separations of 20\,mas--1.2$\arcsec$, reaching contrasts $\Delta m \sim$5--6\,mag at separations around 0.1$\arcsec$. 
We identify 11 (6 previously unreported) binaries having separations in the 10-1000\,au range, and $\Delta m$ between 0.8-5\,mag in our sample. Using archival data to search for components outside our visibility range, we add further multiples (16), which include three triples, leading to a total of 24 multiple systems. Our findings rule out a multiplicity fraction $>$27\% at the 3$\sigma$ level within the speckle observations separation range and detection limits. Future homogeneous spectroscopic/interferometric observations are essential to probe the inner separations, and along with analysis of available astrometry can cover the entire separation range to characterize the multiplicity fraction, and evolutionary scenario of Be stars. 
\end{abstract}



\section{Introduction}


Classical Be stars (CBe) are non-supergiant B spectral-type stars with Balmer emission lines \citep{rivinius}. The emission lines arise from a gaseous circumstellar decretion disk, thought to have formed via rapid rotation \citep[e.g.][]{klement24}. Be stars are  involved in many exotic astrophysical systems, such as X-ray binary stars (Be stars with a neutron star companion), stripped stars (Be stars with a massive, hot companion that has lost its outer envelope and is He-rich), $\gamma$\,Cas stars (highly variable stars with strong X-ray emission). Characterizing Be stars therefore helps better understand a variety of astrophysical phenomena, however, the formation and evolution of Be stars still poses some open questions. 

\begin{figure*}
\centering
\includegraphics[width=1\textwidth]{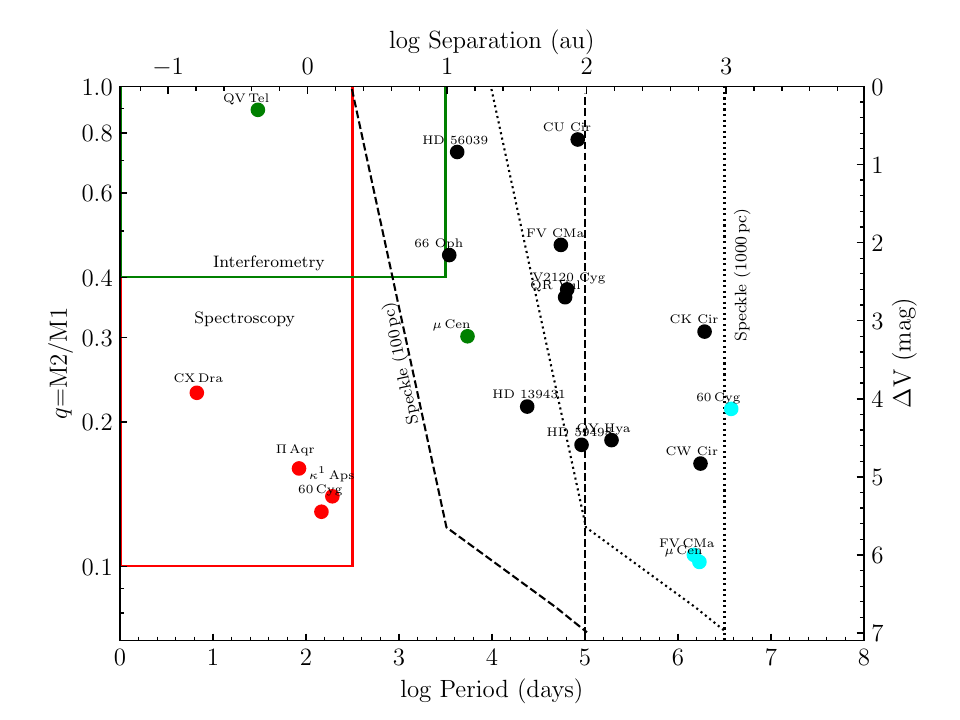}
\caption{The parameter space of detection limits for various methods, adapted from \protect\cite{hutter}. The approximate conversion between the mass ratio, and $\Delta V$ were computed assuming a B0V primary from \protect\cite{pecaut}. The separation and period relation is computed for a B2V/B5V binary, with no eccentricity. For spectroscopy detection limits, we adopt \protect\cite{sana09} where the binary detection probability drops beyond a year for their simulations of massive stars. The dashed and dotted lines represent the speckle detection limits (20\,mas--1.2$\arcsec$) at distances of 100, and 1000\,pc (approx. distances of stars in this study).
The complementarity of different methods in discovering binaries is showcased here. Also shown are detected binaries in our sample from speckle imaging presented in this work (black circles), literature spectroscopy (red circles), archival seeing-limited imaging observations (cyan circles) and interferometric (green) observations. Note that the limits shown here are illustrative, and depend on the exact instrumental configuration for each case.} 
\label{methods}
\end{figure*}

In the past decades, there have been multiple studies constraining the binary statistics of Be stars using either high-resolution imaging \citep{abt84,mason97,oudmaijer, horch, hutter, klementchara, dodd, 2020AJ....159..132S, 2025AJ....169..251G}, SED analysis \citep{klement21}, or compilations of high-resolution spectroscopic data \citep{bod20, abtspec}, and searches for post-interaction binary products such as runaways \citep{boubert, berger}. These studies canvass an important space in the separation/period region of Be stars. But, are limited to probing only close binaries using spectroscopy, or more distant ones using classical seeing-limited imaging techniques (see Fig.\,\ref{methods} for illustrative limits). Speckle imaging allows to search for companions located at angular separations not possible via these methods.

\begin{figure}
\centering
\includegraphics[width=0.5\textwidth]{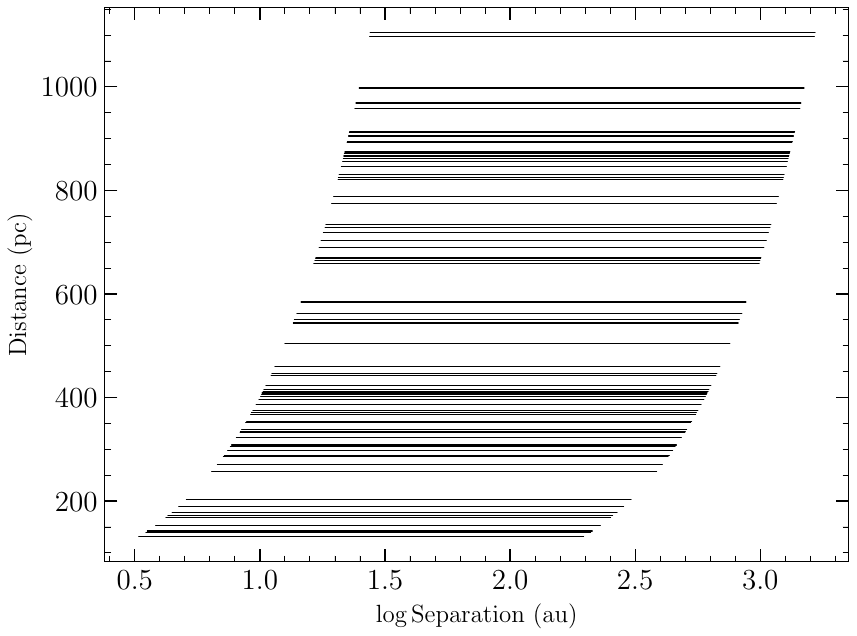}
\caption{The upper and lower limits of separation range from the primary captured by the speckle observations, as a function of distance. Distances are taken from {\it Gaia} EDR3 data of \protect\citet{bailerjones21}. }\label{sep}
\end{figure}

In this paper, we attempt to constrain the multiplicity fraction, and properties of companions of known CBe stars within the local volume of 1\,kpc homogeneously, at binary separations between few to 1000\,au (Fig.\,\ref{sep}), depending on the distance of the source. This corresponds approximately to periods between a few years to a few thousand years for equal mass, early B-type binaries with circular orbits. 
Our data comes from speckle imaging obtained with either the `Alopeke instrument on the Gemini North, or Zorro on the Gemini South twin 8.1\, telescopes which allow for a uniform, homogeneous sample with characterized biases allowing for a statistical inference of our results. Our work complements multiplicity fraction estimates from both spectroscopy (which typically probes much smaller separations), classical seeing-limited imaging or astrometry (at larger separations), and interferometry (which is usually limited to very bright magnitudes, $V\lesssim$8). 

Our paper is organized thus-- in section 2 we present the data used in this study while discussing the sample selection, and biases. Section 3 contains our results on the detected multiplicity fraction, literature cross-matches, and nature of companions. Finally, in Section 4 a discussion on the implications of our results within the current literature and favored Be formation scenario is presented.

\section{Data}

\subsection{Observations and data reduction}

Our input sample is based on known CBe stars located within 1\,kpc. We selected all CBe stars found in the BeSS database \footnote{http://basebe.obspm.fr/}, which contains 2381 such objects (out of a total 2455 Be stars; 8 stars were marked as Classical/Herbig and not considered). Of these, we selected only those stars having known spectral types between B0-B5 (1265), and cross-matched them with {\it Gaia}\,DR2 \citep{dr2} using a radius of 0.3$\arcsec$ (1101), which was the latest catalog available at the time of observation preparation. Stars having parallaxes (accounting for the zero-point offset of 29\,$\mu$as; \citealt{lind}) larger than 1 mas (i.e. located at a distance smaller than 1000\,pc) were selected (341). Three stars located in the Magellanic Clouds were removed (we assumed that {\it Gaia} DR2 parallaxes are erroneous here). Since CBe stars within the local volume of 1\,kpc are selected, no magnitude criteria were applied. The final database consisted of 338 CBe stars which formed our observational sample. All selection catalogs are available from the author on request.  

\begin{figure}
\centering
\includegraphics[width=0.5\textwidth]{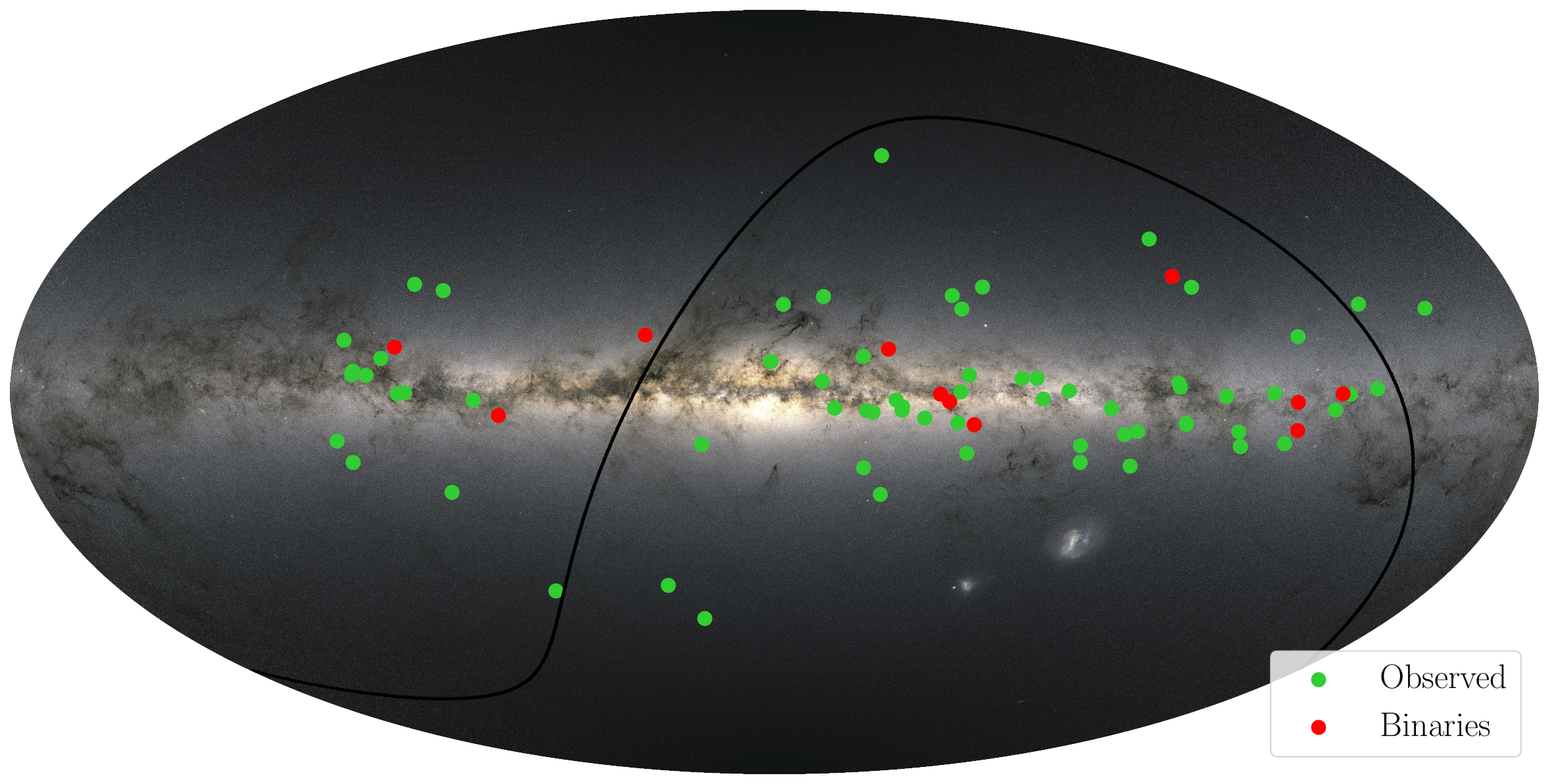}
\caption{Projection of observed targets on all-sky DSS image. Speckle identified binaries are marked.}\label{sky}
\end{figure}

\begin{figure}
\centering
\includegraphics[width=0.45\textwidth]{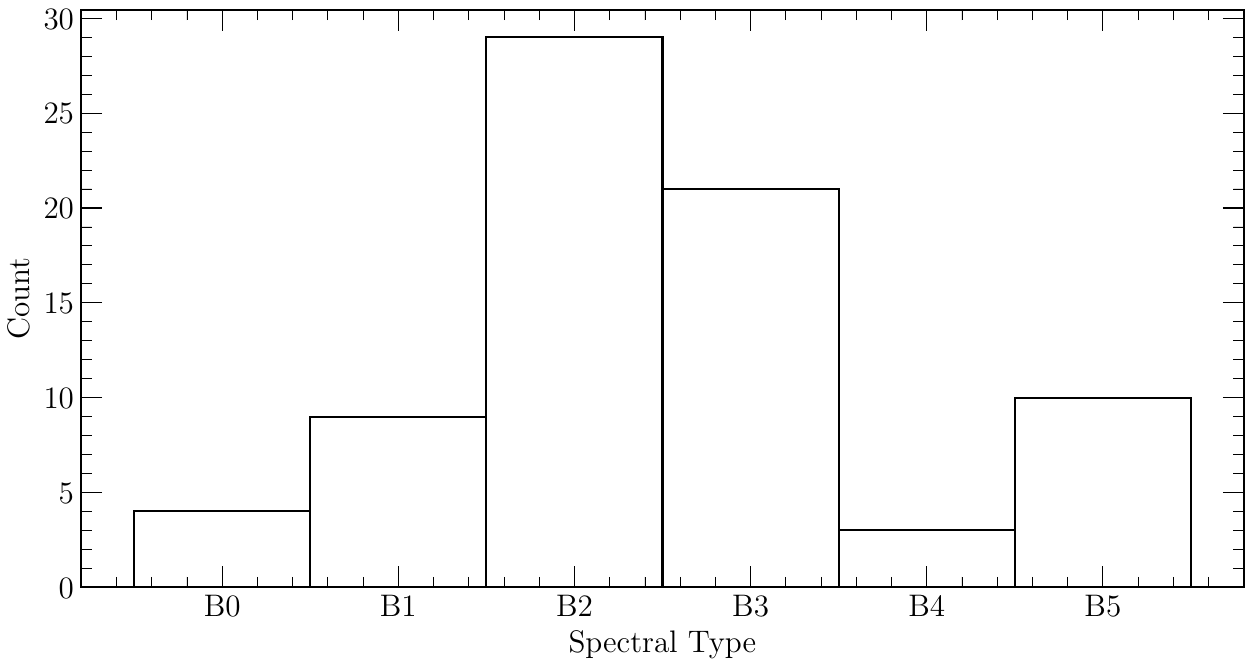}
\caption{Histogram of spectral types for all observed targets.}\label{sptypes}
\end{figure}

Data for our targets were collected between March 2020 and September 2021, using the twin speckle imagers Zorro and $`$Alopeke, mounted on the Gemini South, and Gemini North 8.1m telescopes, respectively \citep{zorro}, and cover the whole sky (see Fig.\,\ref{sky}). 
The observations are taken in custom medium-band speckle filters in a blue and red channel simultaneously (separated at 674\,nm by a dichroic). They are centered at either 466 (EO466) and 716\,nm (EO716), or at 562 (EO562) and 832\,nm (EO832). 
Observations were taken under zenith seeing of less than 0.7'' in clear skies, though with varying moon phases throughout the period. The observations were spread across a two-year period, but not all of the initially selected targets could be observed due to difficulties in scheduling, weather constraints, and the pandemic. The final list of observed targets, along with their known magnitudes and other relevant identifications, is provided in Appendix\,A. These data represent a subset of the initially selected targets, reflecting the observational challenges and scheduling limitations encountered during the campaign. In total, 76 CBe targets are studied here, representing $\sim$21\% of the known CBe stars within a 1\,kpc volume. Their spectral types are given in Fig.\,\ref{sptypes}.

Observations taken at Zorro between March--September 2021, were only taken in the 832\,nm filter due to an issue with blue camera. For these targets, only red speckle imaging is available, which cannot be used for color comparisons of any companion, or estimating stellar properties. All data were processed using the \cite{howell11} pipeline. The pipeline was first developed by \cite{horch01} in which the main Fourier analysis is discussed. Using methods and discussions highlighted in \cite{toko10}, \cite{horch11} and \cite{howell11} provide further details of the methods and the data products which result from the pipeline.

There has not been detected any loss of resolution in the blue channel with the usual narrow band filters we use, as all observations occur at high elevations on purpose. There may be a small loss in a SDSS/broad band filter in the blue, but we rarely use such filters.

We describe briefly the process here. During the reduction, the power spectrum of each image is calculated, and then is corrected for the speckle transfer function by dividing the mean power spectrum of the target by that of the standard star. The pipeline also produces reconstructed images of each target. Fourier analysis is used to identify any multiples in the co-added power spectrum, from any detected fringes. If identified, a fit is used to estimate the angular separation, position angle and magnitude difference. The achieved angular resolutions reached the diffraction limit of the 8.1m telescope. The angular resolutions for the filters EO466, EO532, EO716, and E832 are 15, 17, 22, and 25\,mas, respectively. The contrast ($\Delta m$) limits for each target were determined.

\begin{figure}
\centering
\includegraphics[width=0.5\textwidth]{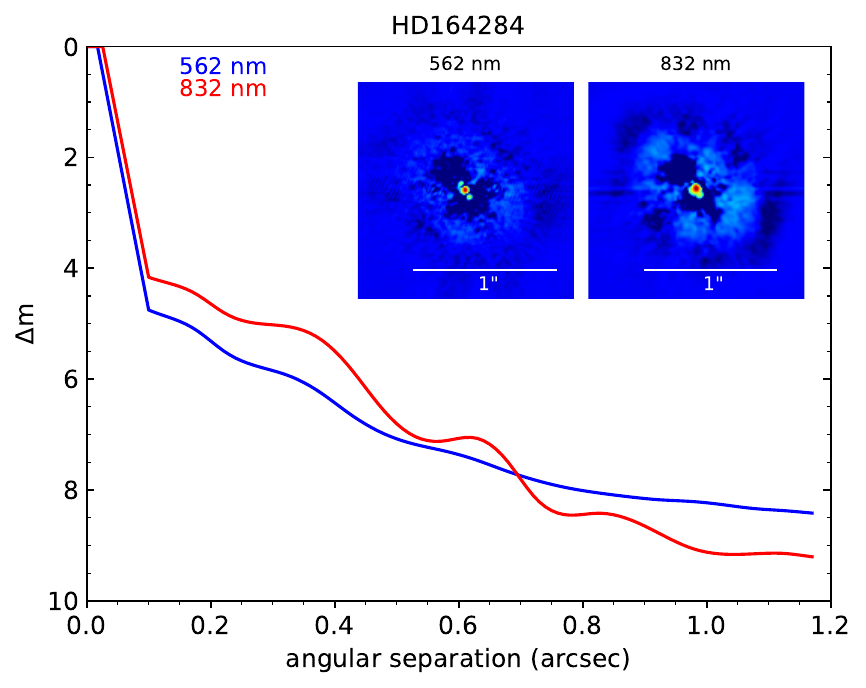}
\caption{Reconstructed image of 66\,Oph in the speckle EO832 filter. The contrast curves for that filter, and the bluer EO562 filter are shown. The companion is visible close to the binary, around 50\,mas away, at a position angle of 211$^\circ$ (a 180$^\circ$ ambiguity exists in the position angle of the companion).}\label{example}
\end{figure}

The method used to compute the contrast curves is described in detail in \cite{horch11}. Here we give a brief overview. The curves are computed in the filters observed by examining the minimum and maximum background values in annuli centered on the primary star. The contrast curves then dictate the observational limit for detecting close companions in relative magnitude compared to the primary star magnitude, as a function of angular separation. A representative reconstructed image, and the corresponding contrast curve of 66\,Oph is shown as an example in Fig.\,\ref{example}. The spline fit to the 5$\sigma$ contrast curve starts with a forced linear segment from the diffraction limit and $\Delta m$=0, to the 5$\sigma$ background fit at 0.1$\arcsec$. This is not a realistic inner contrast limit, but is adopted for spline fitting. To see a realistic set of detections, for example showing that the inner contrast curves do reach the refraction limit, see Fig. 3 in \cite{2021AJ....162...75L}, where there are companion detections ``inside" the spline fit along the diffraction limit. Additionally, using multiple close-in companion detections, the true contrast curve between this region was shown to reach the diffraction limit by \cite{howell25}. Finally, there has not been detected any loss of resolution in the blue channel with the usual narrow band filters used for these observations, and all observations occur at high elevations on purpose as detailed in Appendix A. While there may be a small loss in a broad-band filter in the blue, we do not use such filters for our observations.

All datasets in the raw format are available publicly from the Gemini archive\footnote{https://archive.gemini.edu}, and in the reduced format on the NASA-ExoFOP webpage\footnote{https://exofop.ipac.caltech.edu/}, which includes all reconstructed images, contrast curves, and multiple properties (if present).

\subsection{Effect of runaway stars in the sample}

Finally, we also checked for potential runaways in our sample. We used \cite{gaianss} astrometry to compute the peculiar tangential velocities ($v_t$)$_{\rm pec}$ of our sample stars using the recipe and reference values of \cite{kalarirun}. We find no stars meeting previously used threshold of runaways adopted for early type stars by \cite{moffat} of 42\,kms$^{-1}$. Using a more relaxed criteria of $>$30\,kms$^{-1}$ from \cite{1974RMxAA...1..211C} for peculiar radial velocities ($v_r$)$_{\rm pec}$, we find one star, GP\,Vir. GP\,Vir exhibit's ($v_t$)$_{\rm pec}$=30\,\,kms$^{-1}$; and has a measured ($v_r$)$_{\rm pec}$ from \cite{gaianss} radial velocity of 39\,kms$^{-1}$. We report it as the only runaway candidate in our sample, and suggest the effect of runaways on reported multiplicity statistics of our sample computed using available data is negligible.  


\subsection{Sample incompleteness}

\begin{figure}
\centering
\includegraphics[width=0.45\textwidth]{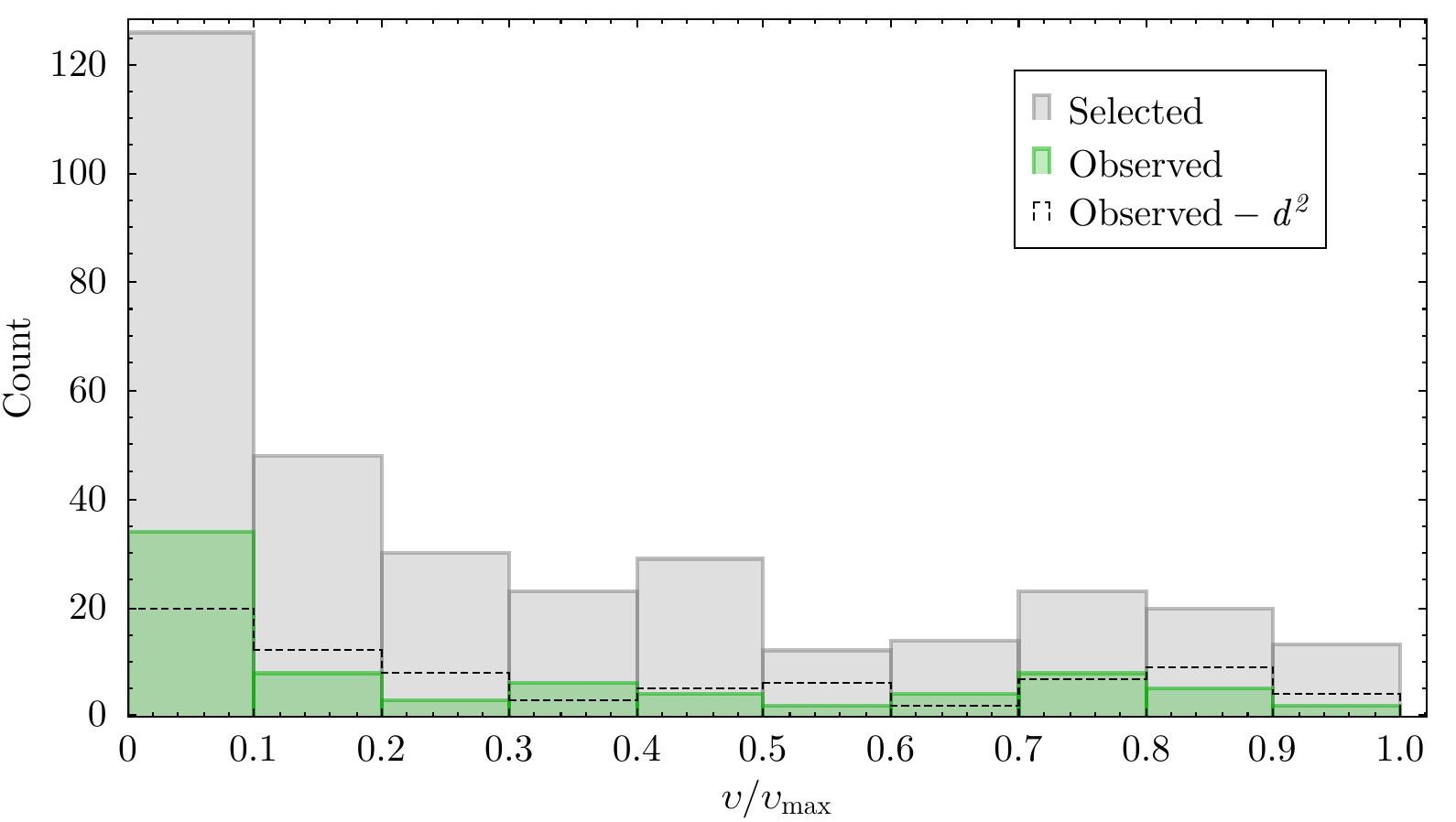}
\caption{$\upsilon/\upsilon_{\rm max}$ of all the selected CBe stars with spectral types between B0-B5 having {\it Gaia} DR2 parallaxes $>$1\,mas found in the BeSS database, and also the observed sample. The dotted histogram represents the $d^2/d^2_{\max}$ values for the observed sample.}\label{vmax}
\end{figure}

Our sample, while volume-limited, is not magnitude limited since most Be stars are within the instrumental detection limit (all stars selected to be observed have $V<$12\,mag). As a test of our sample incompleteness, we compute the ratio of the volume of a given object with respect to the maximum volume, ${\upsilon/\upsilon_{\rm max}}$ \citep{vmax} using the {\it Gaia} DR2 parallaxes adopted for target selection. Given the thickness of the Galactic disc, the numbers are expected to increase as the square of the distance ($d^2$) beyond $\sim$100\,pc, and this is also shown for the sample. For a uniform distribution, the mean of this value should be close to 0.5, however for our sample this is around 0.25. This can be visualized in the distribution of volume of our sources (Fig.\,\ref{vmax}), which are clustered closer towards us for both the observed and target samples.

We interpret this as a lack of distant Be stars in our sample, but also as a lack of known Be stars outside the solar neighborhood. This may suggest that the vast majority of Be stars beyond the solar neighborhood ($\gtrsim$\,100\,pc) remain uncatalogued, and that future studies to detect them homogeneously (for e.g. using their emission lines using methods described in \citealt{vioque, ngc6383}, or infrared excesses as shown by \citealt{chen16}) may be necessary to see if these are to be found. Such studies are essential precursors for future statistical analyses regarding CBe stars.  

\subsection{Effect of parallax cuts}

We note that using {\it Gaia} DR2 parallaxes might bias against resolved binaries, since they may not always have {\it Gaia} DR2 astrometry \citep{dr2}. To estimate the impact of this on our final sample, we inspected the catalog of 164 B0-B5 spectral types not having {\it Gaia} DR2 parallaxes. We applied a magnitude cut of $V<12$\,mag (which is the faintest magnitude in our selected sample) giving us 71 stars. Out of these, the majority ($\sim$60) are located along the Galactic plane in known open associations, particularly $\chi$\,Persei, and Carina that are beyond 1\,kpc. They appear most likely Be star members of clusters that are beyond 1\,kpc. We then inspected archival parallaxes from the SIMBAD database\footnote{https://simbad.u-strasbg.fr/simbad/sim-fbasic}, and found only 6 stars with  parallaxes $>$1\,mas (which were also not close to aforementioned clusters), of which four are very bright $V<3$\,mag ($\gamma$\,Cas, $\delta$\,Sco, $\zeta$\,Tau, $\eta$\,Cen) hence not having {\it Gaia} data, and the remaining two are HD\,75925, HD\,72067. Out of these, HD\,72067 and $\delta$\,Sco have close binaries from the Washington Double star catalog \citep{wds}. 
Overall, we note that two known close binaries, and six potential binaries may have been missed because of our {\it Gaia} DR2 parallax criteria, and the effect is not statistically significant. 


\section{Analysis}

\subsection{Speckle binaries}
We have obtained speckle imaging reaching around 20\,mas with contrasts ($\Delta m$) between 1-5\,mag of 76 Be stars tabulated in Appendix\,A. Our results are described below, with a discussion on specific objects given in Appendix\,B.

Of our 76 targets, 11 displayed evidence for a companion using speckle imaging. No higher order multiples were detected. The properties of the binary companions are given in Table\,\ref{bintable}. To estimate the chance of spurious contaminants, we follow \cite{2006A&A...459..909C, pomohaci}, where the chance of spurious contaminants ($P_{\textrm c}$) is given by $1-e^{-\pi d^2\rho}$, where $d$ is the angular separation, and $\rho$ the background source density. $\rho$ is computed using {\it Gaia} DR3 photometry assuming a circle of 1 arcmin$^2$ centered on the primary, with the magnitude limit set to $G<18$\,mag. None of the sources had $P_{\textrm c}$$>$5\%, with all less than 1\%, except CW\,Cir (HD\,134958) and CK\,Cir (HD\,128293) at 1.2 and 3.8\% respectively. We thus conclude that none of the detected binaries are chance superpositions in our sample. In Table\,\ref{bintable}, we give the angular separation in arcsec, and based on the {\it Gaia} DR3 distance \citep{bailerjones21} the separation in au. The reported position angle is given, although some stars may have a 180$^\circ$ ambiguity (see \citealt{howell11}). The $\Delta m$ is given in the EO\,832 filter unless specified, with a detailed explanation of each binary found in Appendix\,B. All of the binaries are within 1000\,au. These are all detached binaries based on their separation (with periods greater than a few thousand days), i.e. not directly interacting.

\subsubsection{Monte Carlo simulations}

Following \cite{kalari24}, we estimate using Monte Carlo simulations the masses of potentially undetected companions, using the tool described in \cite{molusc}. For each source, we extracted the contrast curve in the filter data was taken (only for the red camera) in, and adopted a mass following the spectral type-mass relationship given in \cite{pecaut}. The period and mass ratio distribution of early-type stars is not as well characterized as low-mass stars due to observational limitations \citep{molusc}. We adopt a log-normal period distribution, with a peak at 1000\,days, and slope for the mass ratio distribution, $\gamma$ of $-$1.7 for mass ratios, $q>$0.3, and $\sigma$ of 2.28 following the constraints found for early-type B stars by \cite{distefano}. Uniform orbital inclination, and eccentricity was assumed. For each source, 5 million companions were generated, and the magnitude was computed using the \cite{mesa} stellar models. 

\begin{figure}
\centering
\includegraphics[width=0.5\textwidth]{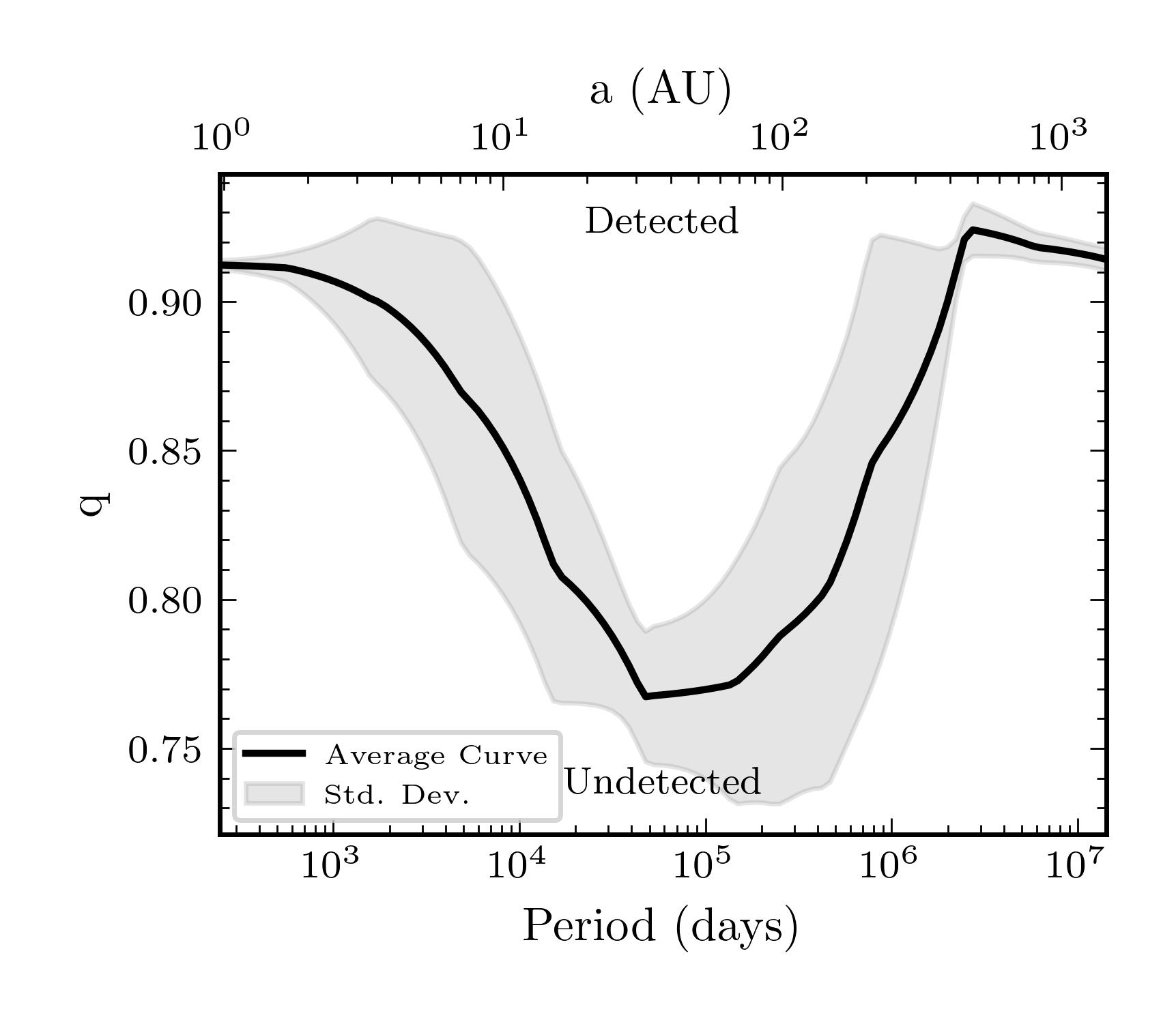}
\includegraphics[width=0.5\textwidth]{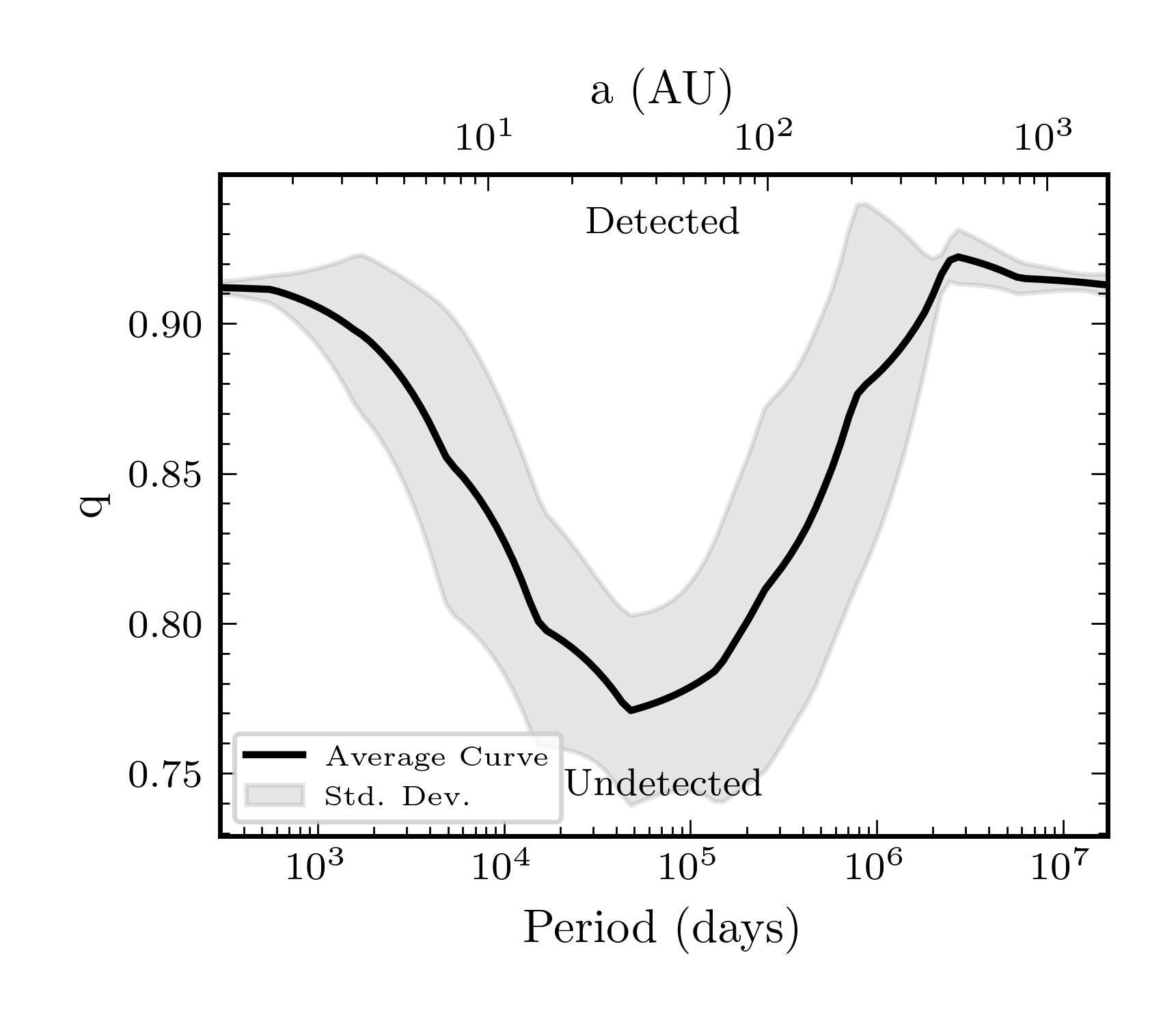}
\caption{Averaged detection limits for companions around Be stars computed using the contrast curves, and converted to mass ratios assuming \cite{mesa} stellar models for the EO\,716 (top) and EO\,832 filters (bottom). Gray shaded regions describe the variation among the detection limits. }\label{molres}
\end{figure}

In Fig.\,\ref{molres} the resulting average detection probability in each filter, along with the spread is shown. The 3$\sigma$ detection probability limits are shown for a given companion mass as a function of the orbital period and separation. As this is estimated using the speckle contrast curve, under the estimated detection limit is the parameter space where a companion is unlikely to be detected using our speckle imaging. This shows via alternative means the discovery parameter space of speckle imaging observed in Fig.\,\ref{methods}. Companions between 10 to a few 100\,au can be recovered by speckle imaging for this sample to around mass ratios of 0.8, while closer-in or further out companions are missed. Low mass ratio close-in binaries have a low probability of detection using speckle, and we cannot statistically rule out binaries in that space using our current observations, but can rule out with high confidence any binaries within $\sim$50--200\,au having $q>$0.8 for the vast majority of our sample, that have been undetected. The complete set of accompanying recovery fractions for each object can be obtained from the principal author on request. 

\subsection{Comparison with literature}

The limitation of our observations are both in separation ranges (20\,mas to $\sim1.2\arcsec$), and in contrast ratios, as depicted in Fig.\,\ref{methods}. The resulting multiple fraction, and detected multiple fall within these limits. However, other observational methods allow for detecting close-in (spectroscopy/interferometry), or further (classical seeing-limited imaging, astrometry) sources. Here, we consolidate our detected binaries with available archival information based on classical, or inteferometric imaging and spectroscopy.

\begin{deluxetable*}{llllllll}
\tablenum{1}
\tablecaption{Properties of multiples detected in our study}
\tablewidth{0pt}
\tablehead{
  \colhead{Name} &
  \colhead{Sp. type} &
  \colhead{$\Delta m^1$} &
  \colhead{Sep.} & \colhead{Sep.} &\colhead{P.A.} &\colhead{Dist.$^2$} & \colhead{Epoch}\\ 
 \colhead{}& \colhead{} & \colhead{(mag)} & \colhead{(au)} &  \colhead{($\arcsec$)} & \colhead{($^{\circ}$)} & \colhead{(pc)} & \colhead{(MJD)}
}
\startdata
  FV\,CMa & B2Vnne & 2.03 & 65.4 & 0.089 & 198.6 & 734.5 & 59227.251748\\
  HD\,56039 & B5Ve & 0.84$^4$ & 11.8 & 0.021 & 193.2$^3$ & 562.3 & 59274.041806\\
  HD\,59498 & B5IVe & 4.59 & 91.9 & 0.105 & 94.7 & 875.5 & 59272.093981\\
  OY\,Hya & B5Ve & 4.53 & 150.8 & 0.495 & 337.7 & 304.6 & 59227.311725\\
CK\,Cir & B2Vne & 3.14 & 700.6 & 1.045 & 295.9 & 670.4 & 59418.068738\\
CU\,Cir & B3Vne & 0.68 & 86.3 & 0.13 & 264.0 & 664.2 & 59418.080498\\
CW\,Cir & B0.5Vne & 4.83 & 655.1 & 0.732 & 162.5 & 894.9 & 59418.091181\\
  HD\,139431 & B2Vne & 4.1 & 37.5 & 0.057 & 292.5 & 658.0 & 59419.073831\\
    66\,Oph & B2Ve & 2.16 & 10.4 & 0.051 & 211.0$^3$ & 203.1 & 59391.442627\\
  QR\,Vul & B3Ve & 2.7 & 70.1 & 0.393 & 192.3 & 178.4 & 59393.509965\\
  V2120\,Cyg & B2Ve & 2.6 & 72.5 & 0.083 & 171.2 & 874.0 & 59389.517546\\
\enddata
\tablecomments{$^1$ Given in the 832\,nm filter; $^2$ Distance from {\it Gaia} DR3 results in \cite{bailerjones21}; $^3$ 180$^\circ$ ambiguity in the fit; $^4$ Given in the 562\,nm filter }
\label{bintable}
\end{deluxetable*}

\subsubsection{WDS catalog counterparts}

To assess if we are missing companions beyond our separation range, we compare our results to the WDS (Washington Double star) Catalog from \cite{wds}. Cross-matching with the identifiers to the J2000 epoch, we find 30 unique companions, which are reported in Table\,\ref{wdstable}. For close separations ($<$100\,mas), we found six sources. For two (66\,Oph and FV\,CMa), the companions were detected in our speckle imaging. The other companions were to 60 Cyg (HIP\,103732), a known multiple star from \cite{koub00}, with the secondary identified as a hot subdwarf from ultraviolet spectra in  \citep{wang17}. The reported companion here is the same, but detected via interferometry by \cite{klement22}. A companion to V4024\,Sgr (a $\gamma$\,Cas variable) was identified, that is marked as a potential binary in \cite{wds}, based on lunar occultation observations of \cite{evans1981}. However, on further inspection, \cite{wang18} using spectroscopy found no companion, and suggested that previous changes reported in the cross-correlation function are due to spectral features of the Be star, and not due to a companion. We therefore discard the notion that V4024\,Sgr is a binary. Interferometric observations of $\mu$\,Cen by \cite{2013MNRAS.436.1694R} detect a companion at 0.1$\arcsec$), and it is marked as a multiple in our study. This companion is at the speckle detection limits, and is not identified in our data. QV\,Tel \citep{frost22} has a companion at 1\,mas detected via interferometric imaging, and predicted via spectroscopy \citep{bodqvtel}. Four other companions are within our detection limits ($\lesssim$1.2$\arcsec$), and all (companions to QR\,Vul, OY\,Hya, CK\,Cir), but one (FV\,CMa) are found in previous speckle imaging with similar separations and $\Delta m$. FV\,CMa has a closer companion detected previously in speckle imaging by \cite{2012AJ....143...42H} and our images, but this companion (found by \cite{oudmaijer} at a separation of 0.7$\arcsec$, $\Delta m\sim$6) is undetected from the speckle analysis, possibly due to the combination of the secondary's brightness, and the tertiary being close to the detection limit (and also in the infrared). It is reported as a triple in Table\,\ref{imagetable}.

\begin{deluxetable*}{lllllllll}
\tablenum{2}
\tablecaption{WDS multiples cross-matches}
\tablewidth{0pt}
\tablehead{
  \colhead{Name} &
  \colhead{WDS} &
  \colhead{Sep.} &
  \colhead{P.A.} &
  \colhead{$V$} &
  \colhead{$\Delta m$} &
  \colhead{$\rho$} &
  \colhead{Probability$^1$}   \\
  \colhead{} &
  \colhead{Identifier} &
  \colhead{($\arcsec$)} &
  \colhead{$^{\circ}$} &
  \colhead{(mag)} &
  \colhead{(mag)} &
  \colhead{(sources/$\arcmin^{{2}}$)} &
  \colhead{(\%)} &
  } 
\startdata
  FV\,CMa$^2$ & 07074-2350 & 0.09 & 201 & 5.71 & 1.99 & 15 & 0.01\\
  FV\,CMa$^3$ & 07074-2350 & 0.7 & 228 & 5.7 & 6.0$^5$ & 10 & 0.43\\
  NV\,Pup & 07183-3644 & 241.6 & 102 & 4.66 & 0.41 & 4 & 100.0\\
  NW\,Pup & 07183-3644 & 118.9 & 215 & 5.07 & 3.6 & 4 & 100.0\\
  $o$\,Pup & 07481-2556 & 26.9 & 197 & 4.5 & 8.1 & 59 & 100.0\\
  I\,Hya & 09413-2335 & 51.8 & 293 & 4.77 & 6.19 & 2 & 99.08\\
  OY\,Hya$^2$ & 09591-2357 & 0.5 & 341 & 6.15 & 4.29 & 1 & 0.02\\
  $\mu$\,Cen$^3$ & 13496-4228 & 0.1 & 80 & 3.5 & 3.2 & 8 & 0.01\\
    $\mu$\,Cen$^3$ & 13496-4228 & 4.6 & 304 & 3.97 & 6.09$^5$ & 8 & 13.73\\
        $\mu$\,Cen & 13496-4228 & 45.5 & 127 & 3.46 & 9.5 & 10 & 100.0\\
  V795\,Cen & 14150-5705 & 5.3 & 296 & 4.83 & 10.24 & 36 & 58.62\\
  V795\,Cen & 14150-5705 & 37.8 & 235 & 5.03 & 7.47 & 36 & 100.0\\
  V795\,Cen & 14150-5705 & 32.3 & 165 & 5.03 & 5.97 & 36 & 100.0\\
  CK\,Cir & 14395-6812 & 16.2 & 256 & 6.76 & 6.08 & 25 & 99.67\\
  CK\,Cir$^2$ & 14395-6812 & 1.1 & 295 & 6.91 & 3.15 & 25 & 2.61\\
  $\kappa^1$\,Aps$^4$ & 15315-7323 & 27.4 & 255 & 5.49 & 5.78 & 6 & 98.04\\
  MQ\,TrA & 16037-6030 & 53.0 & 179 & 7.13 & 0.98 & 14 & 100.0\\
  66\,Oph$^2$ & 18003+0422 & 0.05 & 216 & 5.0 & 1.5 & 4 & 0.0\\
  QV\,Tel$^3$ & 18171-5601 & 0.02 & 137 & 5.9 & 0.3 & 5 & 0.0 \\
  $\lambda$\,Pav & 18522-6211 & 60.6 & 205 & 4.22 & 8.18 & 4 & 100.0\\
  V4024\,Sgr$^4$ & 19083-1917 & $<$0.1 & $-$1 & 5.5 & 3.6 & 7 & 0.01\\
  QR\,Vul$^2$ & 20153+2536 & 0.4 & 190 & 4.8 & 2.75 & 11 & 0.15\\
  QR\,Vul & 20153+2536 & 115.7 & 83 & 4.8 & 4.9 & 11 & 100.0\\
  V2120\,Cyg & 20255+5441 & 47.3 & 147 & 7.25 & 3.95 & 10 & 100.0\\
  V2120\,Cyg & 20255+5441 & 50.3 & 204 & 7.25 & 4.55 & 10 & 100.0\\
  60\,Cyg$^3$ & 21012+4609 & 2.9 & 159 & 5.4 & 4.13 & 22 & 14.91\\
  60\,Cyg$^3$ & 21012+4609 & 0.04 & 14 & 6.7 & $-$- & 22 & 0.0\\
  $\epsilon$\,Cap & 21371-1928 & 65.8 & 46 & 4.49 & 5.62 & 1 & 97.71\\
  $\epsilon$\,Cap & 21371-1928 & 62.7 & 164 & 4.49 & 9.61 & 1 & 96.76\\
  V423\,Lac$^3$ & 22558+4334 & 28.8 & 167 & 8.0 & 1.54 & 3 & 88.6\\
  \enddata
\tablecomments{$^1$ Refers to the chance alignment probability. $^2$Identified in speckle imaging $^3$Added as archival multiple $^4$Rejected as binary, see Section 3.2.1. $^5$Reported in the $K$-band. }
\label{wdstable}
\end{deluxetable*}

We now inspect the remaining 20 stars for possible companions (all at separations greater than 2\,arcsec), which were detected based on classical seeing-limited imaging. To estimate the likelihood of the companions being related, we compare the {\it Gaia} DR3 reported astrometry (parallax, proper motions), and the chance alignment probability described in Section 3.1. The binaries are listed in Table\,\ref{imagetable}. For the binaries, we find that $\mu$\,Cen has a companion (detected in adaptive optics images of \citealt{oudmaijer}) at 4.3$\arcsec$ with $\Delta m$ of 6\,mag (in $K$). Given the similarity in astrometry (within 1$\sigma$) for the 4.3$\arcsec$ companion, we consider this as a potential wide triple companion. Although the chance alignment probability of V423 Lac is high (90\%), the companion reported in WDS has the same {\it Gaia} parallaxes and proper motions (less than 1$\sigma$ difference) as the primary, suggesting that the binary maybe be physical. We therefore consider it as a candidate companion.  
60\,Cyg also a $\Delta m$=4.13 companion within 2.9$\arcsec$, which has a low chance alignment probability, which we mark a triple to the inner spectroscopic companion given the similarity in astrometry to the primary. We note that although  $\kappa^1$\,Aps has a reported close binary companion (1470\,au) in \cite{lindroos}, the system is likely not physical given the significant differences in parallax and properties, but a chance superimposition instead. 
The remaining sources are much farther out with very high chance alignment probabilities, and have {\it Gaia} astrometry more than 3$\sigma$ different from the primary suggesting a chance alignment and are hence rejected. 

Overall, based on the WDS compilation, we catalog companions from the WDS catalog (given in Table 3) based on archival interferometric imaging for 60\,Cyg (added in Table\,4 as detected spectroscopically as well), QV\,Tel, $\mu$\,Cen, archival infrared AO imaging for FV\,CMa, $\mu$\,Cen, and classical seeing-limited imaging for V423\,Lac and 60\,Cyg. 


\begin{deluxetable}{lllllll}
\tablenum{3}
\tablecaption{Properties of multiples reported based on archival observations listed in WDS}
\tablewidth{0pt}
\tablehead{
  \colhead{Name} &
  \colhead{Sp. type} &
  \colhead{$\Delta m^1$} &
  \colhead{Sep.} & \colhead{Sep.} &\colhead{P.A.} &\colhead{Dist.$^2$} \\ 
 \colhead{}& \colhead{} & \colhead{(mag)} & \colhead{(au)} &  \colhead{($\arcsec$)} & \colhead{($^{\circ}$)} & \colhead{(pc)}
}
\startdata
FV\,CMa$^{5}$ & B2Vnne & 6.0$^{3}$ & 587.6 & 0.7 & 228 & 734.52\\
  $\mu$\,Cen & B2Vnpe & 3.2 & 14.0  & 0.1 & 80 & 139.89 \\
  $\mu$\,Cen & B2Vnpe & 6.09$^3$ & 643.5  & 4.6 & 304 & 139.89 \\
  QV\,Tel & B3IIIpe & 0.3 & 0.44 & 0.001 & 137 & 364\\ 
    60\,Cyg$^4$ & B1Ve & 4.13 & 1087.7 & 2.9 & 159 & 375.06 \\
V2155\,Cyg$^{6, *}$ & B1Ve & $-$- & $-$- & $-$- & $-$- & 1104.5\\
  V423\,Lac$^*$ & B3Vne & 1.54 & 16831.3  & 28.8 & 167 & 584.42 \\
\enddata
\tablecomments{$^1$Reported from WDS catalog in $V$ except where references mentioned; $^2$Distance from {\it Gaia} DR3 results of \cite{bailerjones21} $^3$ $K$-band from \cite{oudmaijer} $^4$ Potential triple to inner spectroscopic companion; $^5$Triple, secondary is detected in speckle imaging at 65\,au. $^6$ Candidate binary based on {\it Gaia} \texttt{RUWE} parameter. $^*$ denotes candidate binaries. }
\label{imagetable}
\end{deluxetable}

\subsubsection{Spectroscopic counterparts}

For each source, we searched for possible spectroscopic counterparts. To do so, we used the listed SIMBAD database object types, where we found 5 classified as potential spectroscopic binary candidates ($\chi$\,Oph, $\pi$\,Aqr, CX\,Dra, EW\,Lac, HD\,134401), with the references tabulated in Table\,\ref{spectable}. The first three are cataloged in the survey of \cite{sb9}. The remaining two are listed from the {\it Gaia} non-single stars catalog identified using {\it Gaia} multi-epoch radial velocity spectra combined with astrometry, and are marked as candidate binaries. 
We also cross-matched with the compilation of bright Be binaries from \cite{bod20} with 11 cross-matches, where in addition to the previously identified binaries (and in some cases, non-detections), we report one possible binary subdwarf candidate, $o$ Pup \citep{koub12}. We found one companion in the catalog of \cite{rivi06}, $\epsilon$\,Cap with a period of 128.5 days. $\kappa^1$\,Aps is a spectroscopic binary, with sdB companion reported in \cite{wang23}. We found no new companions when comparing with the LAMOST spectroscopic double lined survey \citep{lamost}.

Therefore, in addition to our 11 speckle companions, we identify seven companions in the WDS catalog, and nine spectroscopic companions from the literature. A further candidate binary is added based on {\it Gaia} astrometry (V2155\,Cyg; see Section 3.2.3). Of these multiples, three are triples ($\mu$\,Cen, 60\,Cyg, FV\,CMa). These literature companions are listed in Tables\,3 and 4. The catalog of WDS companions, including the determination of the companion parameters is presented in Table\,\ref{wdstable}. 

\begin{deluxetable}{lllllll}
\tablenum{4}
\tablecaption{Properties of multiples detected based on archival spectroscopic data}
\tablewidth{0pt}
\tablehead{
  \colhead{Name} &
  \colhead{Sp. type} &
  \colhead{Period} & \colhead{$q$} &\colhead{Ref.} &\colhead{Dist.$^1$} \\ 
 \colhead{}& \colhead{} & \colhead{(days)} & \colhead{} & \colhead{} & \colhead{(pc)}
}
\startdata
$o$\,Pup & B1IVnne & 28.9 & $-$- & 7 & 354.50 \\
HD\,134401$^*$ & B2Vne & 1113.34 & $-$- & 5 & 968.76 \\
 $\kappa^1$\,Aps & B2Vnpe & 192.1 & 0.14 & 12 & \\
$\chi$ Oph & B2Vne & 138.8 & $-$- & 2 & 153.00 \\
CX\,Dra & B2.5Ve & 6.70 & 0.23 & 4 & 351.23 \\
60\,Cyg$^8$ & B1Ve & 146.6 & 0.13 & 9, 10 & 375.06 \\
 $\epsilon$\,Cap & B3Vpe & 128.5 & $-$- & 11 & 270.56 \\
 $\pi$ Aqr & B1Ve & 84.1 & 0.16 & 3 & 333.05 \\
 EW Lac$^*$ & B3IVpe & 4.56 & 0.99 & 5, 6 & 285.61 \\
\enddata
\tablecomments{$^1$ Distance from \cite{bailerjones21}; $^2$ \cite{abt05}; $^3$ \cite{bjorkman}; $^4$ \cite{cxdra}; $^5$ \cite{gaianss}; $^6$ Candidate in \cite{klementchara} based on SED $^7$ \cite{koub12}; $^8$ Detected both in spectroscopy and inteferometry; $^9$ \cite{koub00}; $^{10}$ \cite{klement22}; $^{11}$ \cite{rivi06}; $^{12}$ \cite{wang23}. $^*$ denotes candidate binaries. }
\label{spectable}
\end{deluxetable}

\begin{figure}
\centering
\includegraphics[width=0.5\textwidth]{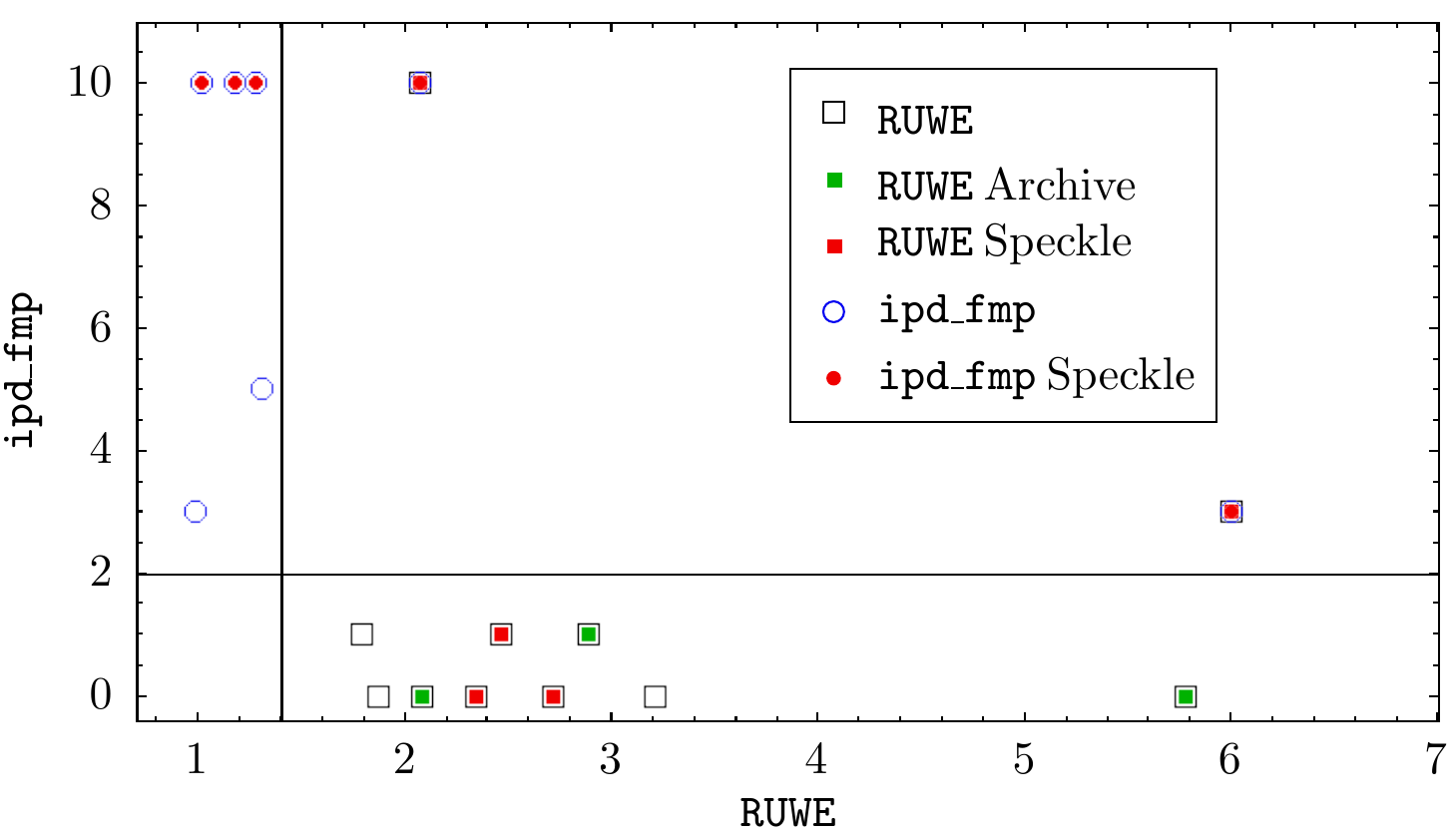}
\caption{Candidate binaries meeting either of the {\it Gaia} binary indicators. $\texttt{RUWE}$ candidates are shown by black squares, with the cut-off of 1.4 marked, and $\texttt{ipd\char`_frac\char`_multi\char`_peak}$ candidates are shown as blue circles where the cut-off ($>$2) is shown by a solid line. Detected archival, or speckle companions are shown by filled green and red markers respectively. For visual purposes, the maximum $\texttt{ipd\char`_frac\char`_multi\char`_peak}$ 
 value was set to 10, and the $\texttt{RUWE}$ value to 6. }\label{gaia}
\end{figure}


\subsubsection{{\it Gaia} multiplicity indicators}

{\it Gaia} provides multiple parameters to identify potential multiples based on astrometry (besides the spectroscopic candidates identified in Section 3.2.1). Here, we consider two primary criteria, which are listed below--
\begin{enumerate}
    \item The Gaia Renormalized Unit Weight Error (termed \texttt{$\mathtt{RUWE}$}) is the root of the normalized $\chi^2$ of the astrometric fit to along-scan observations. For candidates with multiple companions, this values is expected to be $>$1.4 \citep{gaialind}. This parameter can detect companions between 0.04--0.7$\arcsec$ \citep{dodd}. 
    \item The percentage of windows used for astrometric processing that contain more than one peak is sensitive (the {\it Gaia} \texttt{$\mathtt{ipd\char`_frac\char`_multi\char`_peak}$} parameter) to close binaries, as it produces multiple peaks in some scan directions. Following \cite{toko23}, we identify candidates having values greater than 2.
\end{enumerate}

In Fig.\,\ref{gaia}, we compare the first two indicators provided from the {\it Gaia} parameters. Here, we find that all stars meeting both criteria are detected as binaries in our sample. However, three stars meeting the $\texttt{RUWE}$ parameter (NV\,Pup; V2155\,Cyg, and f\,Car) were not identified as binaries in our sample. NV\,Pup was identified as a potential binary (with a separation of 240$\arcsec$ given in Table\,\ref{wdstable}) in our sample. However, if the high $\texttt{RUWE}$ is caused by a closer companion if present, we do not find any literature identification as a binary for NV\,Pup, which is a known $\gamma$\,Cas variable. Similarly, f\,Car is also a variable star \citep{ruban}. Variability is expected to affect the measured \texttt{RUWE} parameter \citep{berklov}. Additionally, although \cite{fitton} have reported that extended discs of asymptotic giant branch stars might inflate the \texttt{RUWE} parameter, this is not strictly applicable to Be stars, which 
have smaller discs. However, even adopting a stricter cut-off $\texttt{RUWE}>$2, we still have V2155\,Cyg as a binary which remains unidentified in our sample. No literature information was found on its binary status. We mark it as a potential binary. Two stars (z\,Pup, and HD\,55135) are found as a potential binary from the peak fraction, but no literature information could be found on their status. These are much lower than the cut-off suggested by other works (for e.g. \citealt{cifuentes} suggest \texttt{$\mathtt{ipd\char`_fmp}$}$>$30, although CU\,Cir has a \texttt{$\mathtt{ipd\char`_fmp}$} of 3, but an $\texttt{RUWE}>$10). We therefore, suggest a cut-off in \texttt{$\mathtt{ipd\char`_fmp}$}$>$10 for Be stars.

We add, from comparison with {\it Gaia} binary indicators one potential binary V2155\,Cyg (but without binary parameters). The remaining candidates with stringent cut-offs (\texttt{$\mathtt{ipd\char`_fmp}$}$>$10; or $\texttt{RUWE}>$2) are all detected in either speckle or in the WDS catalog. We suggest that for Be disc bearing stars, more conservative cut-offs are necessary to identify binarity. 

\section{Discussion}

\subsection{Multiplicity fraction}

In our study, we observed 76 Classical Be stars within 1000\,pc (Appendix\,A) using speckle imaging covering a separation range of 20\,mas--1.2$\arcsec$, reaching $\Delta m<$6\,mag. We found 11 companions, six of which were previously unreported in the literature. We then combined our datasets with archival literature, imaging, and {\it Gaia} binary indicators. We found another 16 companions, three of which were triples. The total number of multiples in our sample is 24, resulting in an observed multiplicity fraction of 32$\pm$5\% (error accounting for candidate multiples), but for our speckle candidates only (i.e. with separation range of 20\,mas--1.2$\arcsec$) is $\sim$15\%. No corrections were made for potentially missing companions based on our detection limitations. Compared to the literature, our fraction is similar, but covers a vaster breadth in separation ranges. A comparison to literature derived multiplicity fractions are given in Table\,\ref{comp}.

 \begin{deluxetable*}{llllll}
\tablenum{5}
\tablecaption{Literature Be star multiplicity fractions}
\tablewidth{0pt}
\tablehead{
  \colhead{Method} &
  \colhead{Scope$^1$} & \colhead{Sp. type} &
  \colhead{Fraction} & \colhead{Size} & \colhead{Reference}} 
  \startdata
Spectroscopy$^2$ & $B<5$\,mag & B2--B5IV & 25\% & 42 & \cite{abtspec}\\
Imaging$^3$ (with select spectra) & $V<6.5$\,mag & B2--B7 & 28\% & 80 & \cite{abt84}\\
Speckle$^4$ & $V<6.5$\,mag & B1--B8 & 10\% & 48 & \cite{mason97}\\
AO imaging$^5$ & $K<7$\,mag & B0--B9 & 30\% & 39 & \cite{oudmaijer}\\
Spectroscopy$^6$ & $V<12$\,mag & B0-B1.5 & 10\% & 287 & \cite{bod20}\\
{Interferometry} & $V<5$ & B0-B9 & 45\% & 31 & \cite{hutter}\\
{\it Gaia}$^7$ & $-$- & B0-B9 & 29\% & 123 & \cite{dodd}\\
{Speckle}$^8$ & $V<11$ & O8-B9 & 26\% & 46 & \cite{2025AJ....169..251G}\\
{Speckle}$^9$ & $<1000$\,pc & B0-B5 & 14\% & 76 & This work\\
{Speckle}, and archival$^{10}$ & $<1000$\,pc & B0-B5 & 32\% & 76 & This work\\
\enddata
\tablecomments{$^1$Magnitude or distance limit; $^2$Around 10 epochs per star, resolution of 5 km\,s$^{-1}$ ; $^3$ from Bright star catalog; $^4$Separations 0.035--1.5$\arcsec$ and $\Delta m<$3.0; $^5$Separations of 0.1--8$\arcsec$, $\Delta m<$10\,mag; $^6$Literature analysis; $^7$Combination of {\it Gaia} PMa; and RUWE parameter, quoted separations of 0.02$\arcsec$--1.1$\arcsec$; $^8$ Separations of of 0.06--9.7$\arcsec$, with $\Delta m <4.8$\,mag $^9$0.020--1.2$\arcsec$, with $\Delta m<$2--6\,mag; $^{10}$ Archival imaging and spectroscopic observations along with speckle imaging}
\label{comp}
\end{deluxetable*}

Following the Clopper-Pearson method described in \cite{kalari24}, we can rule out a multiplicity fraction greater than 47\% within our speckle detection limits (see Fig.\,\ref{molres}), at the 3$\sigma$ confidence level. For speckle only companions, this falls to 27\%. Our speckle multiple fraction is on the lower end of multiplicity studies for Be stars. From speckle interferometry \cite{hutter} found a 45\% multiplicity fraction for nearby bright Be stars, \cite{oudmaijer} reports a Be binary fraction of 30\% from AO imaging, and \cite{dodd} show a 29\% binary fraction from {\it Gaia} proper motion anomaly study. Our values are similar to limited speckle surveys (\citealt{mason97} with separations greater than 0.03$\arcsec$; or \citealt{2025AJ....169..251G} with separations between 0.09--0.33$\arcsec$); or the literature spectroscopic study of \cite{bod20} which probes a smaller separation range than in this study. {\it It is not apparent if the different separation ranges have different multiplicity fractions due to physical effects, or the observed multiplicity fractions at these different separation ranges are purely due to the observational differences between speckle imaging, interferometry, and spectroscopy.} The speckle images are more sensitive to close binaries than literature imaging and {\it Gaia} surveys (but not always in $\Delta m$). Combining archival imaging observations to encompass binaries at larger separations, our multiplicity fraction still remains low ($\sim$24\%), indicating that a simple lack of candidates beyond our detection limits is not the reason for the low multiplicity fraction observed for our targets. Including spectroscopic companions, our multiplicity fraction is around $\sim$32\%, similar to recent literature studies. 

We consider the observed multiplicity fraction a combination of multiple effects. It is most likely that our sample, with a mean distance of around 580\,pc is further than most previous studies. In our case, the angular resolution achieved probes a smaller physical separation range in au (i.e. spectroscopy/interferometry/{\it Gaia} probes closer separation ranges; while imaging finds companions further out). C.f. a median distance of 280\,pc in \cite{dodd}; a limit of $K<6$\,mag in \cite{oudmaijer}; $V<5$\,mag in \cite{hutter} to our mean distance of 580\,pc, which suggests that those studies probe closer multiples ($\sim$5\,au) for the median distance. In addition, our limited separation range, and detection limit means that candidates even if present (for e.g. see FV\,CMa) may not be detected given our observational limitations. 


\subsubsection{Comparison with Proper Motion Anomaly Be binaries}

We compare our catalog to the binaries detected in \cite{dodd}. In that paper, \cite{dodd} used a combination of either {\it Hipparcos} and {\it Gaia} DR2 or DR3 astrometry to compute the proper motion anomaly (PMa; \citealt{kervella}). Essentially the method compares the long-term proper motion vector (measured over the almost 25 years elapsed between the {\it Hipparcos} and {\it Gaia} DR3 data acquisition for example) with the short-term proper motion as measured by either {\it Gaia} or {\it Hipparcos}. For a single star, the proper motions would be similar, in case of a binary system, a change in proper motion (the PMa)  indicates orbital motion in an otherwise unresolved binary system. \cite{dodd}  determine that the {\it Hipparcos --Gaia} DR3 PMa is sensitive to binary systems with separations from about 20 mas to the spatial resolution of {\it Gaia}, 0.7 arcsec. They also found that a magnitude difference of at least 4 can be probed using the method, with as proviso that the smaller the magnitude difference, the smaller any change in motion of the photo-center and thus PMa will be.

Here we concentrate on the {\it Hipparcos}--{\it Gaia} DR3  PMa as that has the best separation overlap with our speckle data. Ten of the 11 binary systems in Table~\ref{bintable} have a listing in the PMa catalogue by \cite{kervella}. Seven (70\%) of these are identified as a binary system based on  their large PMa.  Three objects have a PMa signal-to-noise ratio less than 3, and are thus not recognized as a binary. It may be useful to point out that of these three, CW Cir has a separation of 0.732 arcsec in our speckle imaging, and this is at the higher separation limit probed by the {\it Hipparcos}--{\it Gaia} DR3  PMa, whereas HD 56039 and CU Cir have very small magnitude differences of 0.84 and 0.68 mag. respectively. This will have drastically reduced the PMa values. Hence, the detection statistics of the speckle binary systems are consistent with the PMa.  

Of the 65 objects that are not found as binary in our data, 53 are present in the catalog \cite{kervella}, of which 9 (17\%) have a significant PMa. As the PMa is capable of identifying binary systems with magnitude differences larger than the speckle imaging $\Delta m$ limit, we suspect that these systems have too faint companions to be detected in the speckle data. We suggest that further high-resolution imaging, or spectroscopic follow-up is necessary to verify the Be binaries identified through {\it Gaia} PMa values where possible.

\subsubsection{Limitations}
Although the multiplicity fraction of early-type stars is very high (approaching 100\%), the vast majority are close-in, detected at less than 10\,mas \citep{frost25}. In that paper, 72\% of B stars identified by interferometry have binaries, however, fewer 20\% of binaries detected would have been detected by speckle imaging as they are extremely close ($<25$\,mas). Future follow-up observations in the spectroscopic and interferometric space are necessary to populate the parameter space not covered in this study.

\subsection{Nature of detected companions}

\begin{figure}
\centering
\includegraphics[width=0.5\textwidth]{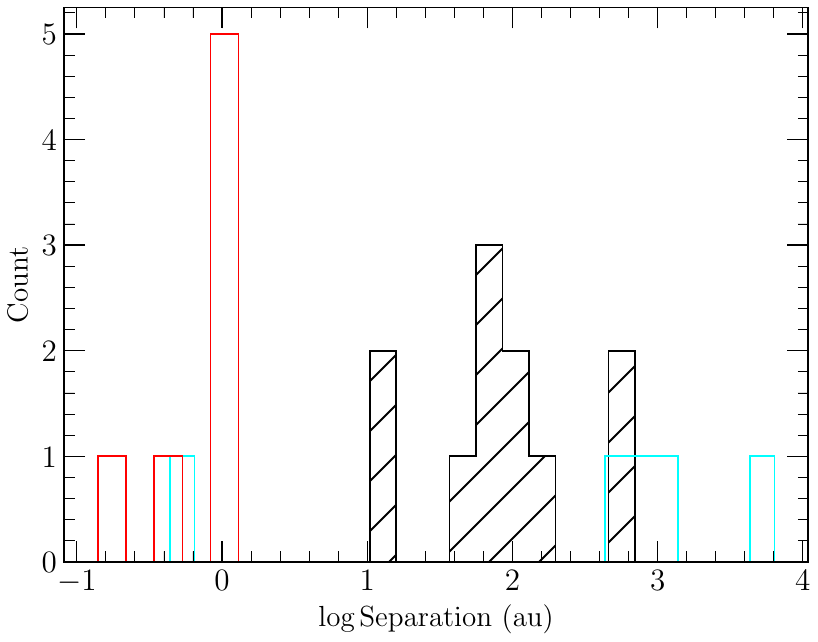}
\caption{Histogram of separations of known companions. No candidates are included. Black bars represent companions identified in speckle imaging, whereas cyan and red bars are literature imaging and spectroscopic companions respectively. The cyan bar at separation less than 1\,au is the interferometric binary identified in QV\,Tel. The information presented is given in tabular form in tables 1, 3, and 4.}\label{aurange}
\end{figure}

We plot the separation range of all companions in Fig.\,\ref{aurange}, and describe the characteristics in Table\,6. 
The build-up of close-in companions are due to the spectroscopic companions, and the interferometrically identified binary of QV\,Tel. For spectroscopic binaries, we estimated the separation based on the period, and the mass of both components from the spectral type following \cite{pecaut}. Where the companion spectral type was unknown, we assumed a mass ratio of 0.5. 

The speckle companions and archival imaging companions have separations $>10$\,au. 
Currently, this result is not in agreement with a flat distribution ({\"O}pik's law) noted in the literature \citep{offner}. There is a build-up of close-in binaries identified spectroscopically, and a lack of binaries beyond this range. We speculate that this result could be a natural consequence of Be star formation via binary interaction, where a close-in companion is necessary to form the Be star \citep{bod20, dodd}. However, there is an important caveat. At these distances ($>$few\,au), we are constrained by the detection limits of our survey, and the additional literature data. For example, we identify few binaries between $>$2,000--10,000\,au, which we consider a result of the instrumental field of view. Similarly, for the speckle separation range (between roughly few\,au--few 100s of au depending on the distance), the probability to detect companions with $q<$0.8 is significantly lower when adopting the period and mass ratio distributions of B stars.  
Therefore, future dedicated spectroscopic studies, combined with analysis of available astrometric data can help populate further any multiples beyond these detection limits, and confirm if Be stars conform to {\"O}pik's law for companion separation. 

For objects with sufficient information to enable companion analysis, we list the nature of the secondary companion, and if known, the orbital period in Table\,\ref{details}. From this we can see that the majority of the speckle companions are early-type main-sequence stars, with exception of OY\,Hya. For spectroscopically identified close binaries, there are potentially few main-sequence targets, for example $\pi$\,Aqr and CX\,Dra. Following \cite{bod20}, similar mass main-sequence companions close enough to interact are not expected to be found for Be stars for the binary evolutionary scenario. They discuss the case of $\pi$\,Aqr, and discard it as the companion is not massive enough. Similarly, we also can discard CX\,Dra as a comparison to their hypothesis, but the candidate companion to EW\,Lac identified in \cite{gaianss} is predicted to be a near equal-mass close-in companion and deserves further follow-up.

\begin{deluxetable}{llll}
\tablenum{6}
\tablecaption{Primary and companion characteristics}
\tablewidth{0pt}
\tablehead{
  \colhead{Name} &
  \colhead{Primary} & \colhead{Secondary} & \colhead{Period} \\
  \colhead{} &
  \colhead{Sp. type} & \colhead{Sp. type} & \colhead{} } 
\startdata
\multicolumn{2}{l}{Spectroscopy}\\
$\kappa^1$\,Aps$^5$ & B2Vnpe & sd0 & 192.1$d$\\
CX\,Dra$^2$ & B2.5V & F5III & 6.7$d$\\
60\,Cyg$^4$ & B1Ve & sd0 & 146.6$d$\\
$\pi$\,Aqr$^1$ & B1Ve & A-FV$^*$ & 84.1$d$\\
EW\,Lac$^3$ & B3IVpe & B3-B4$^*$ & 4.6$d$\\
\multicolumn{2}{l}{Speckle imaging}\\
FV\,CMa & B2Vnne & B8V$^7$ & 162\,$yr^8$\\
OY\,Hya & B5Ve & G$^7$ & 812\,$yr^8$\\
66\,Oph$^6$ & B2V & B8V & 64.2\,$yr$\\
QR\,Vul & B3Ve & A3V$^7$ & 217\,$yr^8$ \\
V2120\,Cyg & B2V & B6V$^7$ & 235.4\,$yr^8$\\
\enddata
\tablecomments{$^1$Naze et al. (2017), secondary nature not well constrained; $^2$Berdyugin et al. (2002), Richards et al. (1999); $^3$ based on {\it Gaia} spectroscopy, SED shows downturn. Mass based on ratio; $^4$Wang et al. 2017; $^5$Wang et al. 2023; $^6$ Hutter et al. 2021 orbital and secondary determination; $^7$Approximate, based on stellar mass tracks assuming same age as primary; $^8$ Lower limit estimate based on orbital separation, and mass of system assuming no eccentricity}
\label{details}
\end{deluxetable}

\subsection{Formation mechanisms}

In this paper, we attempted to constrain the multiplicity fraction of CBes across the few au to few thousand au range. For most of our targets, we cannot rule out close binaries, and thereby Be binaries formed via mass transfer leading to an evolved close-in companions (e.g. see \citealt{klement24}). However, if we consider the assumption that all our stars have a close-in undetected companion, some will be triple or higher order multiple systems when combined with the companions presented here (see \citealt{distefano}).
As an e.g., 60\,Cyg has a close-in evolved sdB companion and a wider companion, or FV\,CMa, $\mu$\,Cen have two wider companions ($>$50\,au). In such hierarchical triple systems with a more distant outer body, the Kozai-Lidov mechanism \citep{kozai} and its interaction with the circumstellar disc must be considered, assuming the inclination of the system meets the criteria (inclination between the two bodies differ by more than 39$^{\circ}$). 

In such circumstances, the triple systems are predicted to lead to oscillation driven outbursts \citep{bekldvout}, or changes in the emission line profiles due to disc tearing \citep{belzldv}. Further study of such hierarchical systems, in particular determining the orbital parameters and inclination may help understand the impact of the Kozai-Lidov mechanism on observables in Be stars.

\section{Conclusions}

In this study we observed via speckle imaging 76 known Classical Be stars, ranging from B0-B5 spectral subtypes, located within $\lesssim$1\,kpc from us. The angular separation range probed is between $\sim$5\,au--1000\,au depending on the distance of the source. Our main results are--
\begin{enumerate}
    \item Identification of 11 companions, of which 6 have no previous literature. Complementary literature search revealed another 16 companions (incl. three triples) indicating 24 multiples.  
    \item We rule out a multiplicity fraction greater than 27\% within the detection limits for speckle inteferometry (between 20\,mas--0.1$\arcsec$ to $\sim$1--5,mag, and 0.1$\arcsec$--1.0$\arcsec$ reaching $\Delta m<$5--6\,mag). Combined with literature, we rule out a multiplicity fraction $>$47\%, but without the homogeneity afforded by the speckle survey.
\end{enumerate}

\begin{acknowledgments}

We thank the anonymous referee for useful comments which helped improve the manuscript. V. M. K. thanks A. Tokovinin for useful comments regarding QV\,Tel, and S. Deveny for providing the intermediate data reduction products for HD 56039. This research has made use of the SIMBAD and VIZIER databases, operated at CDS, Strasbourg, France. This research made of use of the STARLINK software, TOPCAT. This work has made use of the BeSS database, operated at LESIA, Observatoire de Meudon, France: http://basebe.obspm.fr. Observations in the paper made use of the High-Resolution Imaging instruments $'$Alopeke and Zorro. $'$Alopeke and Zorro were funded by the NASA Exoplanet Exploration Program and built at the NASA Ames Research Center by Steve B. Howell, Nic Scott, Elliott P. Horch, and Emmett Quigley. $'$Alopeke and Zorro are mounted on the Gemini North and South telescopes of the international Gemini Observatory, a program of NSF NOIRLab, which is managed by the Association of Universities for Research in Astronomy (AURA) under a cooperative agreement with the U.S. National Science Foundation. on behalf of the Gemini partnership: the U.S. National Science Foundation (United States), National Research Council (Canada), Agencia Nacional de Investigación y Desarrollo (Chile), Ministerio de Ciencia, Tecnología e Innovación (Argentina), Ministério da Ciência, Tecnologia, Inovações e Comunicações (Brazil), and Korea Astronomy and Space Science Institute (Republic of Korea).

 \end{acknowledgments}

%

\vspace{5mm}
\facilities{Gemini-North ($`$Alopeke); Gemini-South (Zorro)}

\appendix{}

\section{Classical Be stars observed. }

\begin{deluxetable*}{lllrllllll}[h!]
\tablecaption{Classical Be stars observed.}
\tablewidth{0pt}
\tablehead{
  \colhead{Name} &
  \colhead{Sp. type} &
  \colhead{Right Ascension$^1$} &
  \colhead{Declination$^1$} &
  \colhead{$V$} & \colhead{HD} &\colhead{Airmass} &\colhead{MJD} &\colhead{0.1$\arcsec$ Limit} &\colhead{1$\arcsec$ Limit}  \\ 
 \colhead{}& \colhead{} & \colhead{(hh:mm:ss)} & \colhead{(dd:mm:ss)} &  \colhead{(mag)} & \colhead{ID} & \colhead{} & \colhead{ } & \colhead{($\Delta m$)} & \colhead{($\Delta m$)}
}
\startdata
  HD 52812 & B3Ve & 07:01:33.61 & $-$27:13:22.60 & 6.93 & 52812 & 1.0 & 59228.187731 & 4.8 & 8.7\\
  19 Mon & B1Ve & 07:02:54.78 & $-$04:14:21.24 & 5.0 & 52918 & 1.16 & 59253.393218 & 5.3 & 8.2\\
  HD 54086 & B5IIIe & 07:06:52.31 & $-$14:41:54.34 & 9.19 & 54086 & 1.04 & 59228.19213 & 4.8 & 8.7\\
  FV CMa & B2Vnne & 07:07:22.59 & $-$23:50:26.59 & 5.83 & 54309 & 1.09 & 59227.251019 & 4.4 & 8.3\\
  HD 55135 & B2.5Ve & 07:11:20.85 & $-$10:25:43.78 & 7.32 & 55135 & 1.1 & 59228.225289 & 4.9 & 8.8\\
  HD 56039 & B5Ve & 07:14:59.91 & $-$11:52:13.47 & 8.28 & 56039 & 1.06 & 59274.039618 & 4.9 & 8.4\\
  NV Pup & B2Ve & 07:18:18.39 & $-$36:44:02.23 & 4.67 & 57150 & 1.02 & 59273.089734 & 4.8 & 8.3\\
  NW Pup & B2IVne & 07:18:38.19 & $-$36:44:33.85 & 5.11 & 57219 & 1.03 & 59273.102512 & 5.1 & 8.0\\
  OT Gem & B2Ve & 07:24:27.65 & +15:31:01.91 & 6.41 & 58050 & 1.03 & 59252.395544 & 4.5 & 8.5\\
  HD 59498 & B5IVe & 07:29:22.78 & $-$21:52:09.18 & 7.79 & 59498 & 1.02 & 59272.092523 & 4.6 & 8.1\\
  V373 Pup & B2Vne & 07:29:27.97 & $-$21:51:31.03 & 7.73 & 59497 & 1.03 & 59272.10544 & 4.7 & 8.0\\
  z Pup & B3Vne & 07:33:51.04 & $-$36:20:18.21 & 5.44 & 60606 & 1.01 & 59271.09316 & 4.1 & 8.5\\
  $o$ Pup & B1IVnne & 07:48:05.17 & $-$25:56:13.81 & 4.49 & 63462 & 1.01 & 59228.221667 & 4.6 & 8.1\\
  BT CMi & B2Vne & 07:57:03.99 & +02:57:03.04 & 7.77 & 65079 & 1.21 & 59228.23456 & 4.4 & 7.4\\
  V374 Car & B2IVnpe & 07:58:50.55 & $-$60:49:28.06 & 5.81 & 66194 & 1.18 & 59228.258542 & 4.5 & 8.8\\
  HD 68468 & B3npshe & 08:12:00.39 & $-$14:10:08.37 & 8.3 & 68468 & 1.04 & 59228.239074 & 4.3 & 9.1\\
  $r$ Pup & B1.5IIIe & 08:13:29.52 & $-$35:53:58.27 & 4.77 & 68980 & 1.01 & 59228.249641 & 4.7 & 8.6\\
  HD 69168 & B2Ve & 08:13:45.65 & $-$46:34:43.27 & 6.48 & 69168 & 1.08 & 59271.157072 & 5.0 & 8.3\\
  HD 69404 & B2Vnne & 08:14:51.24 & $-$46:29:09.21 & 6.44 & 69404 & 1.05 & 59228.254097 & 4.5 & 8.3\\
  $f$ Car & B3Vne & 08:46:42.55 & $-$56:46:11.19 & 4.49 & 75311 & 1.14 & 59227.293356 & 4.7 & 8.8\\
  HD 76985 & B5Vne & 08:56:47.14 & $-$59:31:12.01 & 9.05 & 76985 & 1.15 & 59227.282072 & 4.9 & 8.2\\
  IU Vel & B2.5Vne & 09:00:22.26 & $-$43:10:26.36 & 6.08 & 77320 & 1.03 & 59227.271192 & 4.9 & 8.6\\
  E Car & B2IVe & 09:05:38.38 & $-$70:32:18.60 & 4.65 & 78764 & 1.32 & 59227.288171 & 4.8 & 8.3\\
  I Hya & B5Ve & 09:41:17.01 & $-$23:35:29.45 & 4.76 & 83953 & 1.01 & 59227.264722 & 4.5 & 7.6\\
  V485 Car & B3IIIpshe & 09:41:37.30 & $-$68:30:17.96 & 7.1 & 84375 & 1.27 & 59227.298924 & 5.1 & 8.4\\
  HD 85083 & B5IIIe & 09:47:34.01 & $-$58:11:16.70 & 8.27 & 85083 & 1.13 & 59227.304352 & 5.2 & 8.5\\
  OY Hya & B5Ve & 09:59:06.30 & $-$23:57:02.77 & 6.25 & 86612 & 1.01 & 59227.310995 & 4.3 & 7.7\\
  HD 89884 & B5IIIe & 10:21:59.40 & $-$18:02:04.13 & 7.13 & 89884 & 1.03 & 59274.208252 & 5.0 & 7.7\\
  V353 Car & B2Ve & 11:10:02.34 & $-$60:05:42.49 & 7.74 & 97151 & 1.16 & 58924.201817 & 4.0 & 8.3\\
  HD 103574 & B2Ve & 11:55:21.66 & $-$63:42:12.79 & 7.98 & 103574 & 1.2 & 58924.21559 & 4.5 & 8.2\\
  DK Cru & B2IVne & 12:14:01.77 & $-$59:23:48.83 & 8.81 & 106309 & 1.15 & 58924.25206 & 4.6 & 8.1\\
  39 Cru & B5IIIe & 12:41:56.57 & $-$59:41:08.95 & 4.94 & 110335 & 1.15 & 58923.253032 & 5.1 & 8.2\\
  GP Vir & B3e & 13:35:43.32 & $-$06:09:22.05 & 8.01 & 118246 & 1.09 & 58924.283044 & 5.0 & 8.3\\
  $\mu$ Cen & B2Vnpe & 13:49:36.99 & $-$42:28:25.43 & 3.43 & 120324 & 1.03 & 58923.310556 & 4.5 & 8.0\\
  V774 Cen & B3Vne & 13:53:28.23 & $-$39:03:25.93 & 7.61 & 120958 & 1.02 & 58923.318819 & 4.9 & 7.7\\
  V795 Cen & B4Vne & 14:14:57.14 & $-$57:05:10.05 & 5.07 & 124367 & 1.18 & 59418.026354 & 4.7 & 8.4\\
    \enddata
\end{deluxetable*}

\begin{deluxetable*}{lllrllllll}
\tablewidth{0pt}
\tablehead{}
\startdata
  CK Cir & B2Vne & 14:39:31.66 & $-$68:12:12.19 & 6.93 & 128293 & 1.36 & 59418.06728 & 4.9 & 8.1\\
  V1012 Cen & B3Vne & 14:40:05.48 & $-$59:55:52.88 & 9.08 & 128588 & 1.15 & 58924.322361 & 5.1 & 8.6\\
  CU Cir & B3Vne & 15:07:30.08 & $-$60:46:36.53 & 8.54 & 133495 & 1.25 & 59418.077581 & 4.6 & 7.1\\
  HD 134401 & B2Vne & 15:13:12.16 & $-$65:58:09.03 & 8.98 & 134401 & 1.23 & 58924.356157 & 5.0 & 7.4\\
  CW Cir & B0.5Vne & 15:15:16.17 & $-$58:10:22.37 & 8.19 & 134958 & 1.23 & 59418.088993 & 5.1 & 7.6\\
  $\kappa^1$ Aps & B2Vnpe & 15:31:30.82 & $-$73:23:22.53 & 5.49 & 137387 & 1.37 & 58924.360185 & 3.8 & 7.6\\
  HD 139431 & B2Vne & 15:39:45.65 & $-$42:46:02.71 & 7.34 & 139431 & 1.07 & 59419.070914 & 5.2 & 8.0\\
  V1040 Sco & B2Ve & 15:53:55.86 & $-$23:58:41.15 & 5.4 & 142184 & 1.01 & 59274.403646 & 5.1 & 7.6\\
  MQ TrA & B0Ve & 16:03:44.47 & $-$60:29:54.47 & 7.3 & 143448 & 1.16 & 58923.384861 & 4.7 & 8.0\\
  HD 146463 & B3Vnne & 16:19:14.23 & $-$54:57:42.12 & 8.08 & 146463 & 1.1 & 58924.407164 & 5.3 & 8.2\\
  HD 146596 & B5IVe & 16:19:42.67 & $-$52:46:19.03 & 7.98 & 146596 & 1.09 & 58924.419375 & 4.5 & 8.1\\
  HD 147302 & B2IIIne & 16:24:01.27 & $-$55:27:13.37 & 7.72 & 147302 & 1.11 & 58923.392731 & 4.9 & 7.7\\
  $\chi$ Oph & B2Vne & 16:27:01.43 & $-$18:27:22.49 & 4.43 & 148184 & 1.15 & 59418.129722 & 4.7 & 8.9\\
  V846 Ara & B3Vnpe & 16:56:08.84 & $-$50:40:29.25 & 6.33 & 152478 & 1.17 & 59418.156308 & 4.9 & 8.1\\
  HD 153222 & B1IIe & 17:00:28.69 & $-$49:15:14.91 & 8.91 & 153222 & 1.18 & 59418.166481 & 5.3 & 7.5\\
  HD 154218 & B3Vne & 17:05:42.96 & $-$36:44:25.90 & 7.57 & 154218 & 1.09 & 59418.149352 & 5.1 & 7.6\\
  HD 156831 & B3Vnne & 17:20:42.59 & $-$24:16:16.65 & 8.87 & 156831 & 1.06 & 59419.023056 & 5.2 & 8.0\\
  HD 157099 & B3Vne & 17:23:13.57 & $-$42:49:45.21 & 8.83 & 157099 & 1.03 & 58923.420984 & 4.9 & 7.6\\
  66 Oph & B2Ve & 18:00:15.80 & +04:22:07.02 & 4.6 & 164284 & 1.05 & 59391.441169 & 4.2 & 9.1\\
  QV Tel & B3IIIpe & 18:17:07.53 & $-$56:01:24.07 & 5.36 & 167128 & 1.11 & 59419.117789 & 5.8 & 8.3\\
  CX Dra & B2.5Ve & 18:46:43.09 & +52:59:16.66 & 5.9 & 174237 & 1.2 & 59389.444132 & 5.0 & 8.7\\
  $\lambda$ Pav & B2Ve & 18:52:13.03 & $-$62:11:15.33 & 4.21 & 173948 & 1.34 & 59418.265035 & 5.5 & 8.0\\
  HD 175863 & B4Ve & 18:53:44.70 & +60:01:04.33 & 7.03 & 175863 & 1.31 & 59390.467616 & 4.4 & 9.0\\
  V4024 Sgr & B2Ve & 19:08:16.70 & $-$19:17:25.03 & 5.49 & 178175 & 1.02 & 59417.148611 & 5.2 & 8.3\\
  QR Vul & B3Ve & 20:15:15.90 & +25:35:31.05 & 4.75 & 192685 & 1.0 & 59393.508507 & 4.4 & 8.8\\
  V2113 Cyg & B1Vnnpe & 20:16:48.18 & +32:22:47.39 & 7.16 & 193009 & 1.04 & 59391.544734 & 3.8 & 8.5\\
  V2120 Cyg & B2Ve & 20:25:32.81 & +54:41:03.12 & 7.36 & 194883 & 1.22 & 59389.516088 & 4.6 & 8.9\\
  V417 Cep & B1Ve & 20:51:09.99 & +55:29:19.49 & 8.33 & 198895 & 1.23 & 59389.528021 & 4.7 & 9.2\\
  60 Cyg & B1Ve & 21:01:10.93 & +46:09:20.78 & 5.43 & 200310 & 1.12 & 59390.555012 & 4.4 & 9.4\\
  HD 201522 & B0Ve & 21:08:29.63 & +47:15:25.37 & 7.9 & 201522 & 1.13 & 59390.562465 & 4.3 & 8.5\\
  6 Cep & B3IVe & 21:19:22.22 & +64:52:18.68 & 5.18 & 203467 & 1.42 & 59389.579063 & 5.1 & 9.2\\
  V2155 Cyg & B1Ve & 21:24:30.34 & +55:22:00.24 & 7.54 & 204116 & 1.23 & 59391.56265 & 4.5 & 9.4\\
  V432 Cep & B2Vnne & 21:36:59.64 & +58:08:24.61 & 8.54 & 239712 & 1.27 & 59389.572373 & 5.1 & 9.0\\
  $\epsilon$ Cap & B3Vpe & 21:37:04.83 & $-$19:27:57.65 & 4.55 & 205637 & 1.02 & 59416.285498 & 5.0 & 7.7\\
  HD 206773 & B0Vpe & 21:42:24.18 & +57:44:09.80 & 6.87 & 206773 & 1.28 & 59389.601678 & 4.5 & 7.8\\
  16 Peg & B3Ve & 21:53:03.77 & +25:55:30.49 & 5.08 & 208057 & 1.01 & 59391.588831 & 4.7 & 9.3\\
  UU PsA & B4IVne & 22:04:36.77 & $-$26:49:20.50 & 5.95 & 209522 & 1.01 & 59416.308079 & 4.1 & 8.0\\
  $\pi$ Aqr & B1Ve & 22:25:16.62 & +01:22:38.63 & 4.64 & 212571 & 1.34 & 59419.213032 & 5.6 & 7.8\\
  V423 Lac & B3Vne & 22:55:47.06 & +43:33:33.43 & 7.97 & 216851 & 1.1 & 59390.616539 & 4.3 & 8.9\\
  EW Lac & B3IVpe & 22:57:04.50 & +48:41:02.65 & 5.43 & 217050 & 1.15 & 59390.608183 & 4.6 & 9.2\\
\enddata
\tablecomments{$^1$ Given in J2000.}
\end{deluxetable*}

\section{Individual objects}

\subsection{Notes on individual objects}

\begin{figure*}
\centering
\includegraphics[width=0.45\textwidth]{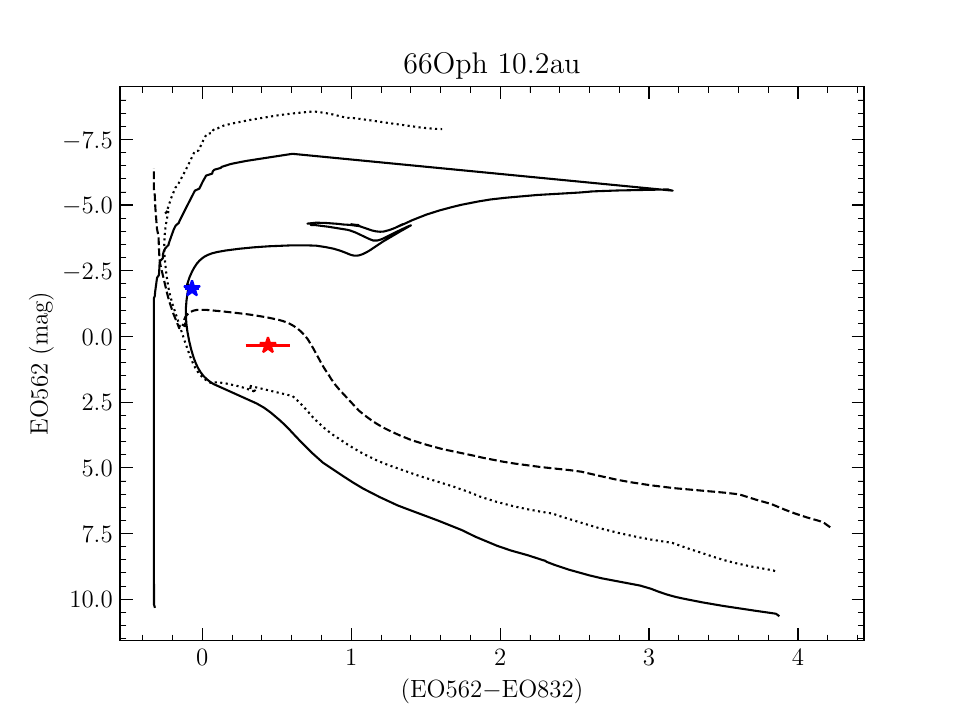}
\includegraphics[width=0.4\textwidth]{TIC404619905I-ef20210626-562_832_plot.pdf}
\includegraphics[width=0.45\textwidth]{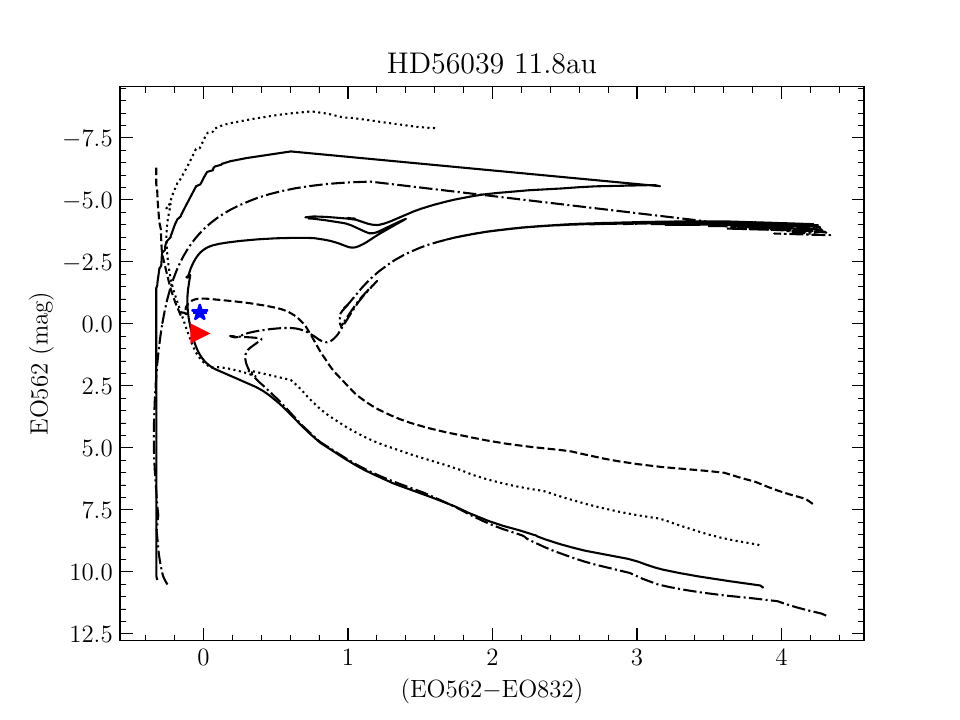}
\includegraphics[width=0.4\textwidth]{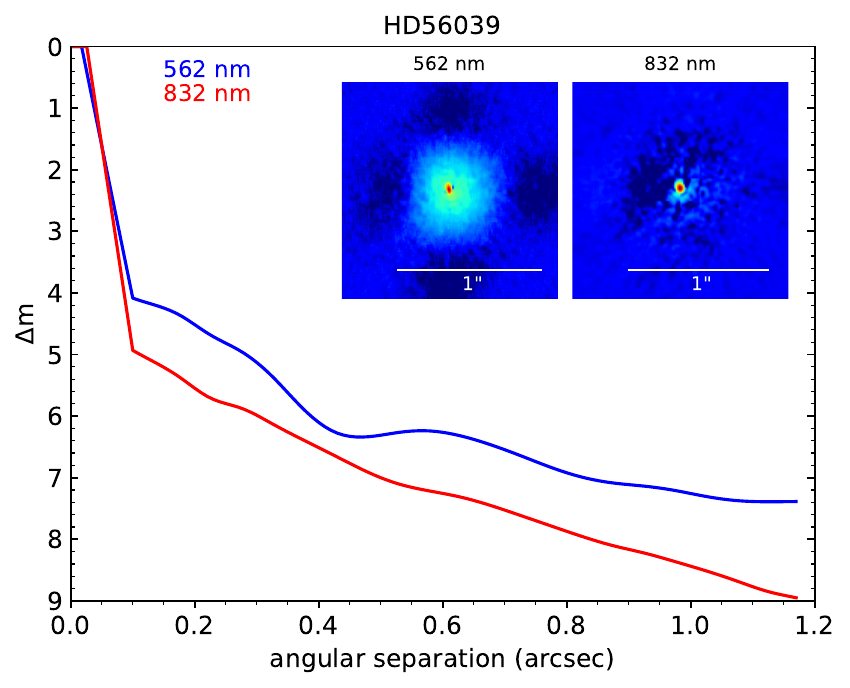}
\includegraphics[width=0.45\textwidth]{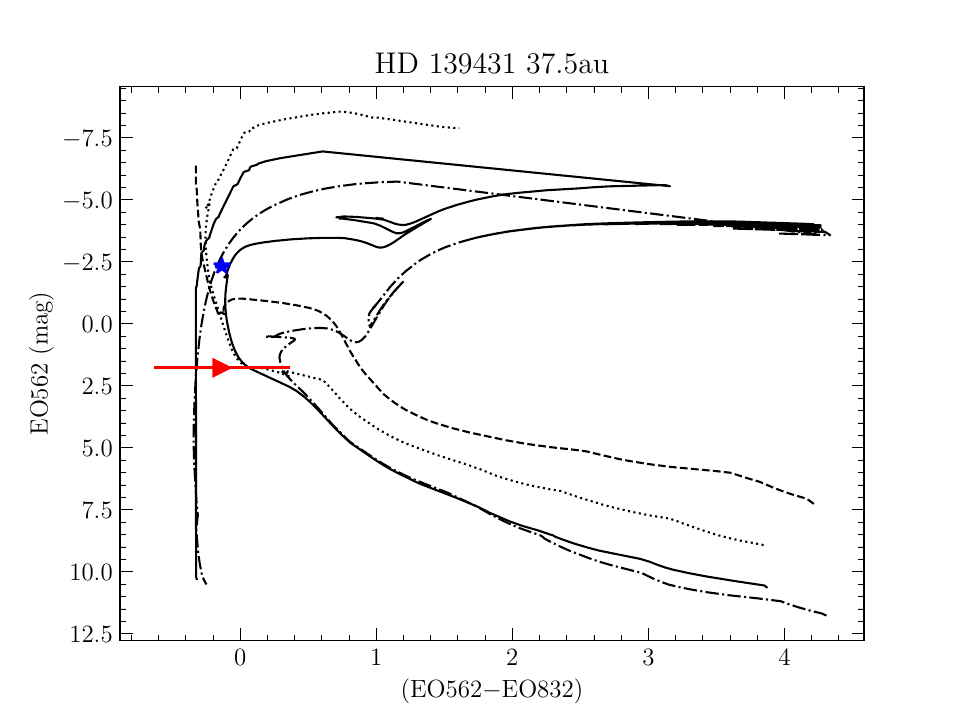}
\includegraphics[width=0.4\textwidth]{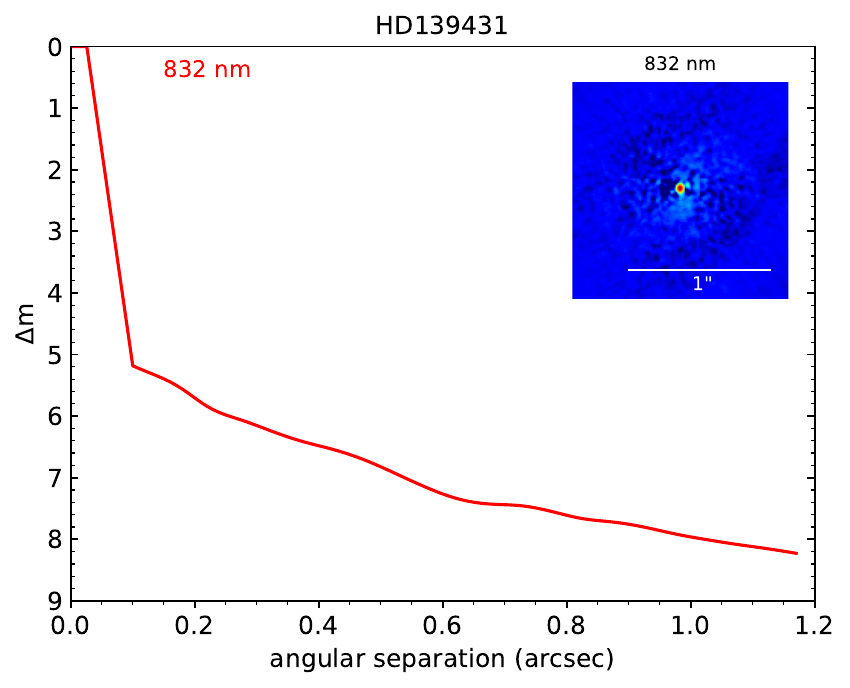}
\includegraphics[width=0.45\textwidth]{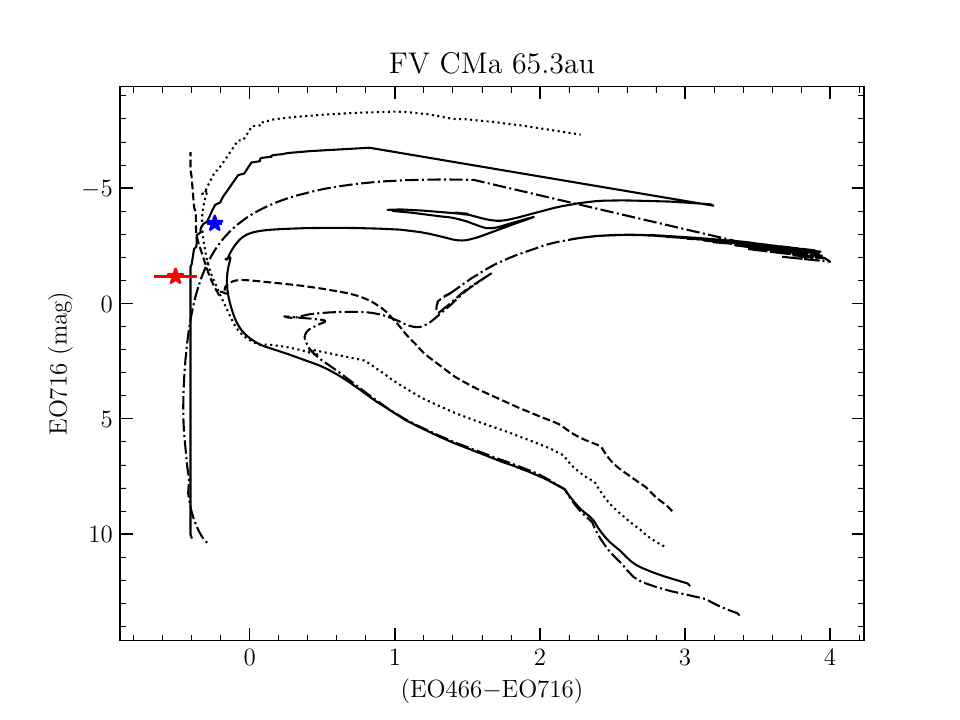}
\includegraphics[width=0.4\textwidth]{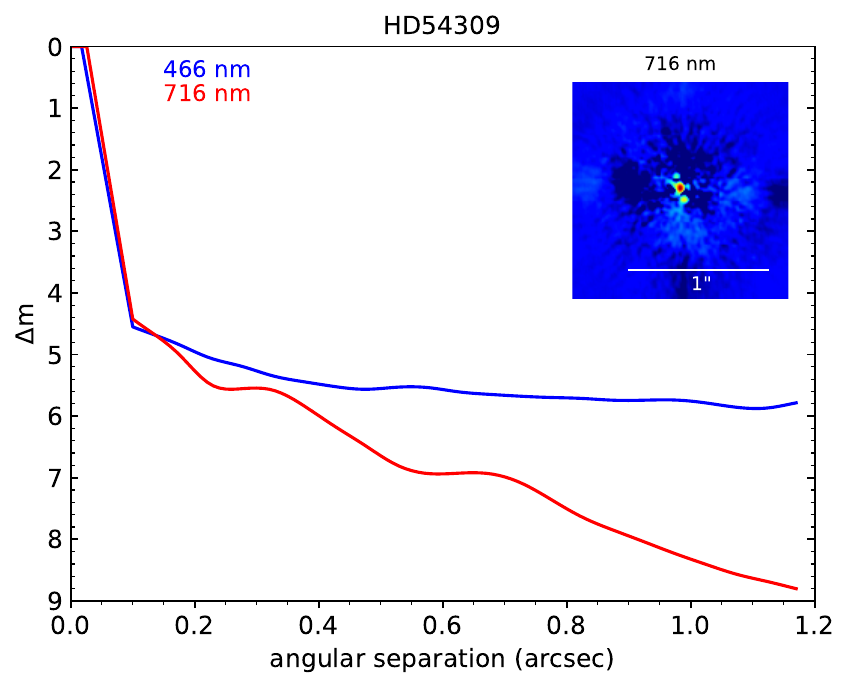}
\end{figure*}
\begin{figure*}
\centering
\includegraphics[width=0.45\textwidth]{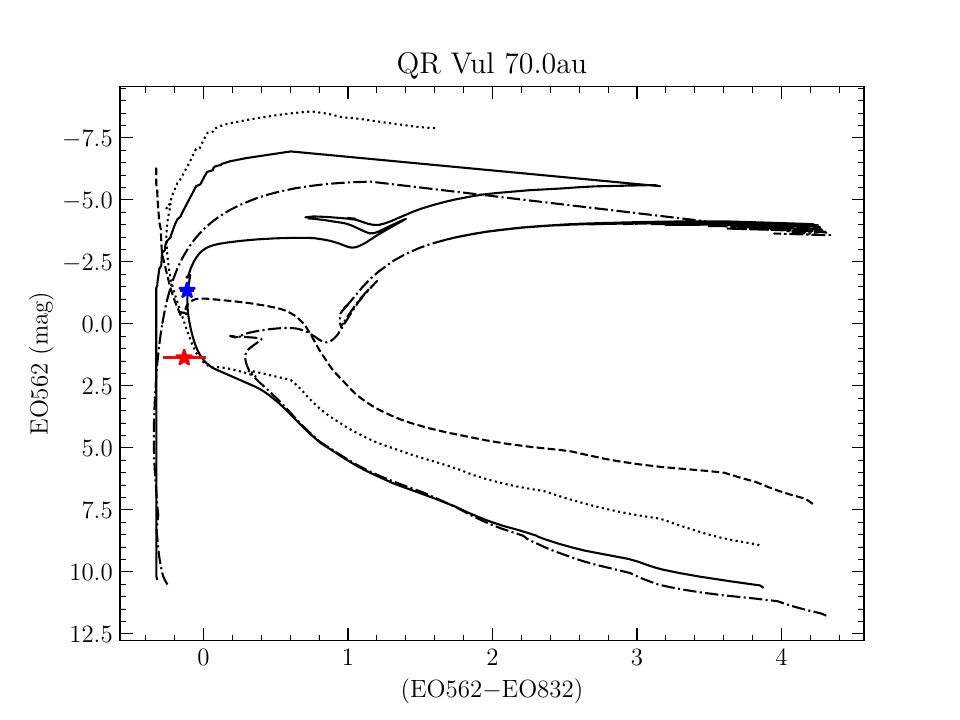}
\includegraphics[width=0.4\textwidth]{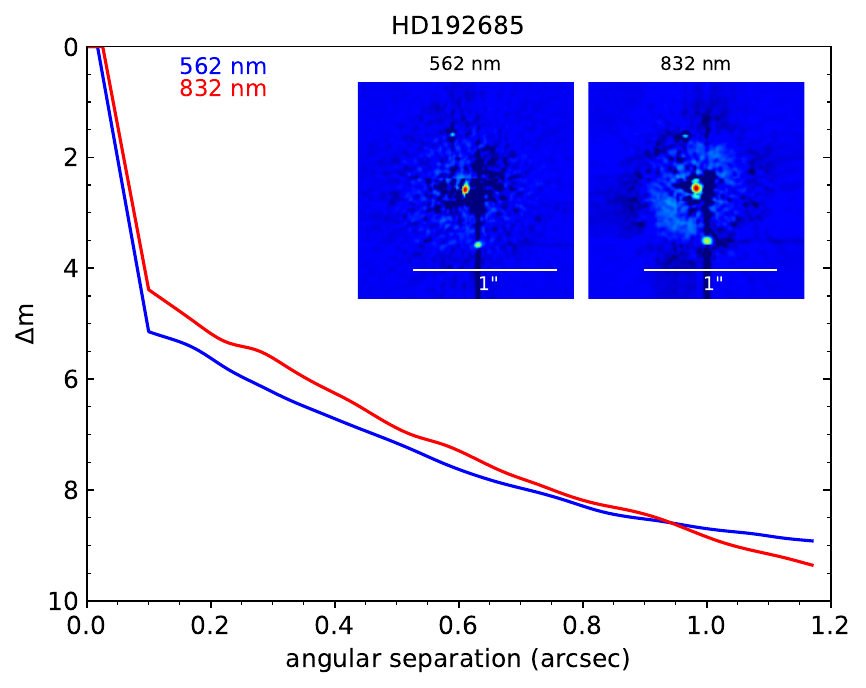}
\includegraphics[width=0.45\textwidth]{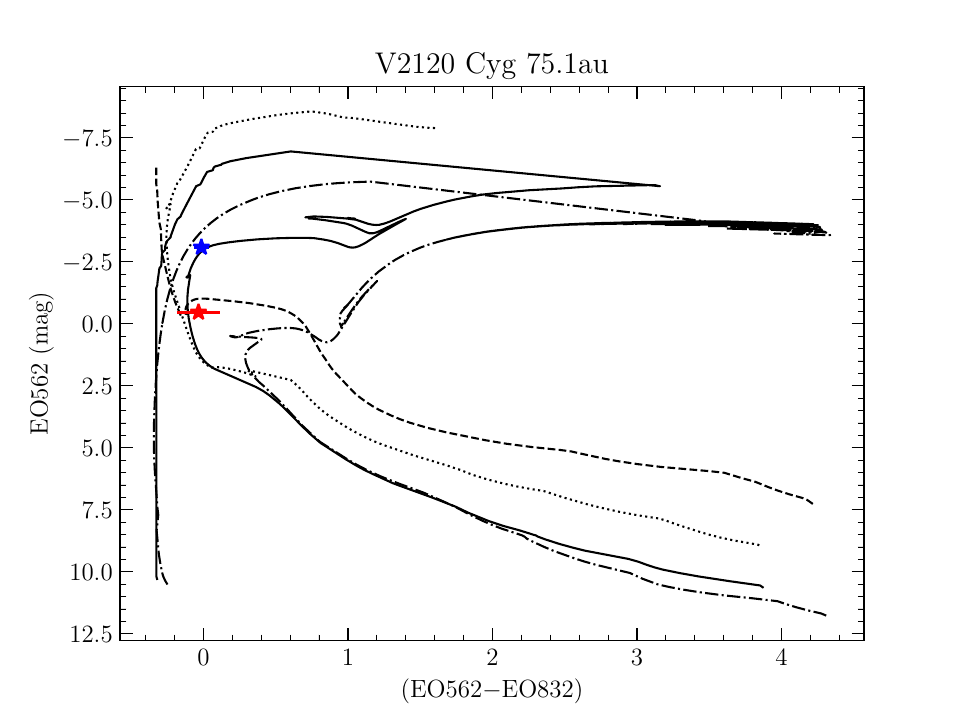}
\includegraphics[width=0.4\textwidth]{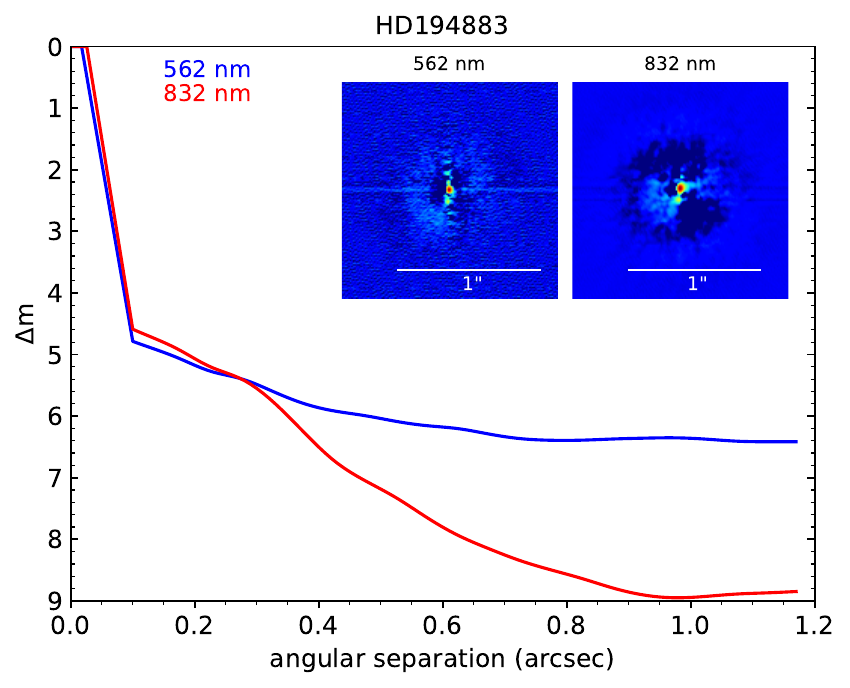}
\includegraphics[width=0.45\textwidth]{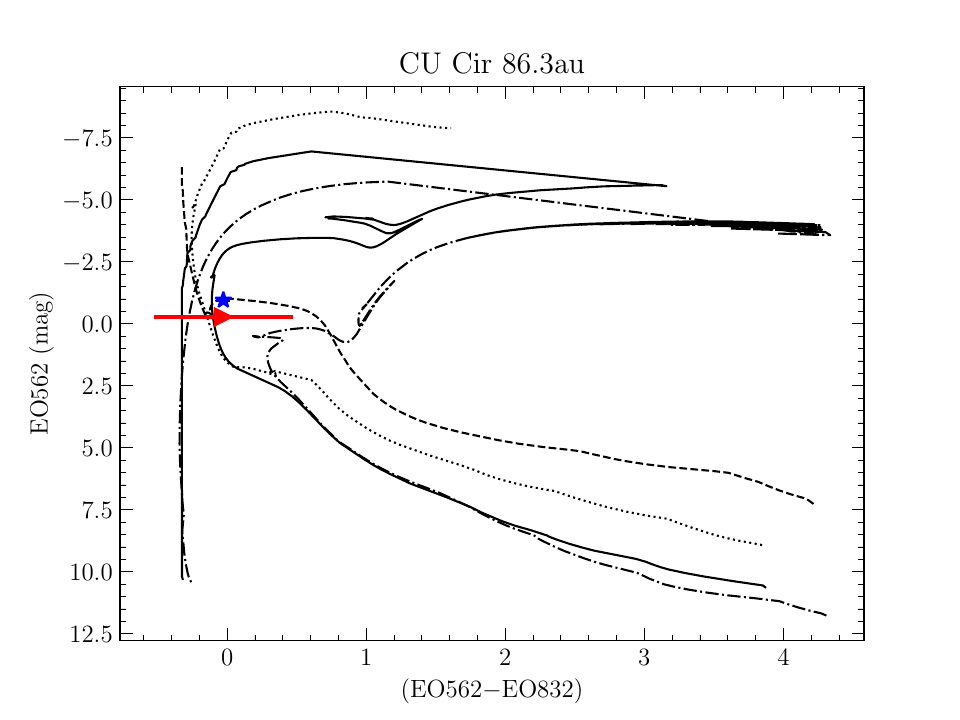}
\includegraphics[width=0.4\textwidth]{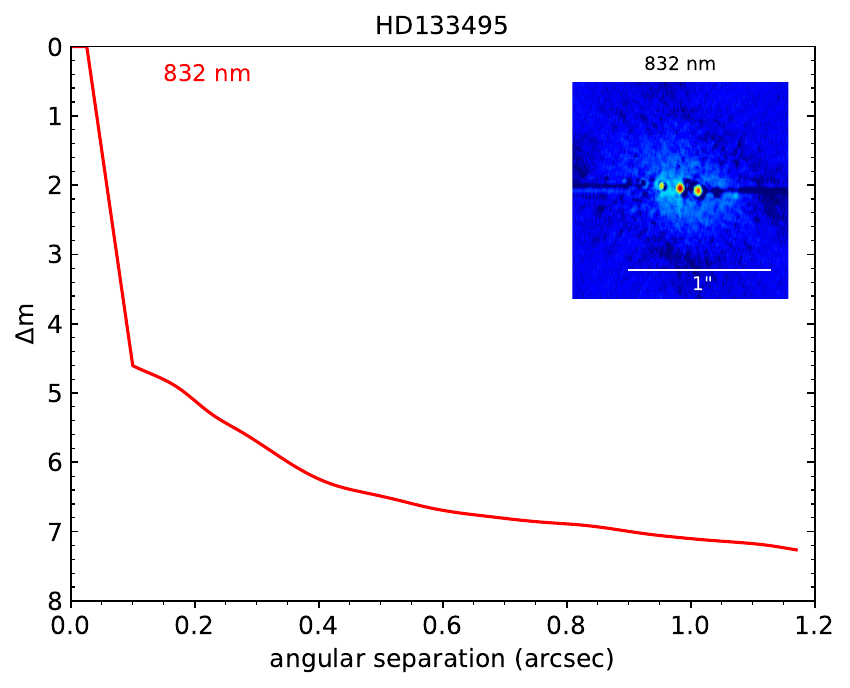}
\includegraphics[width=0.45\textwidth]{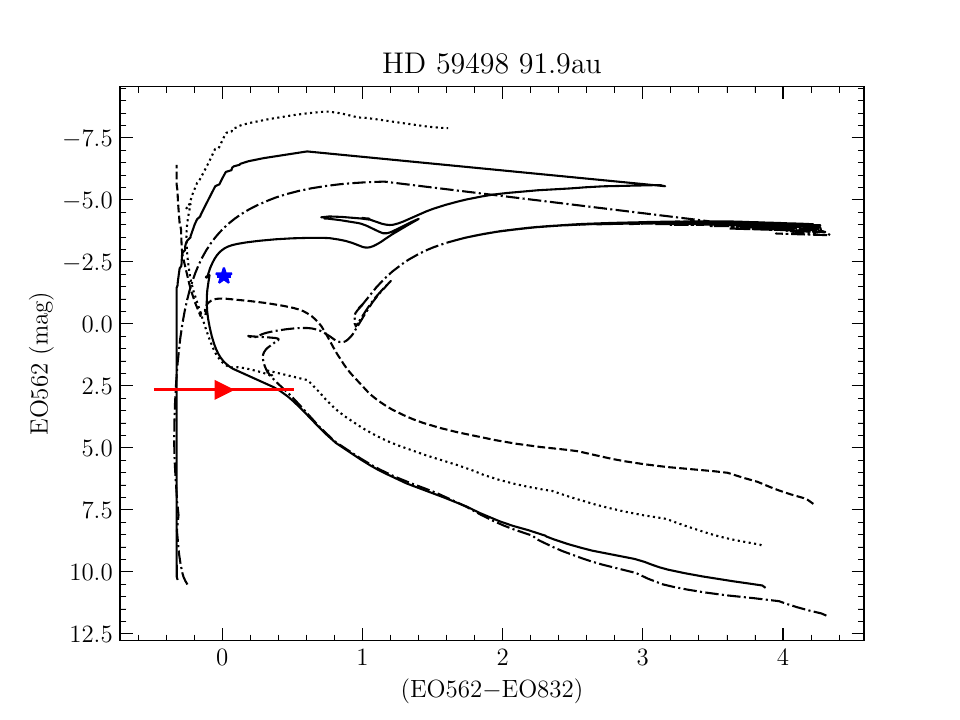}
\includegraphics[width=0.4\textwidth]{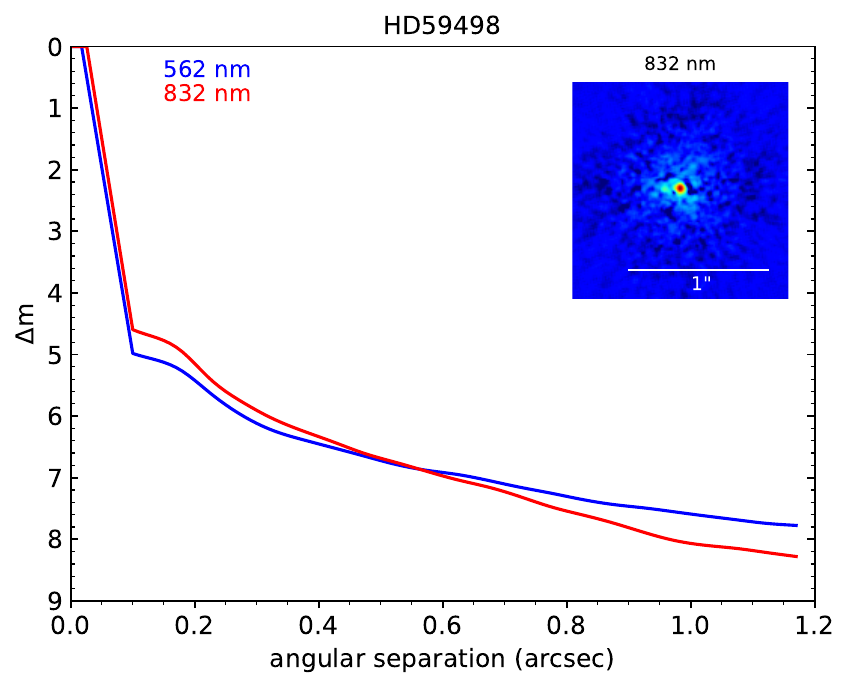}
\end{figure*}
\begin{figure*}
\centering
\includegraphics[width=0.45\textwidth]{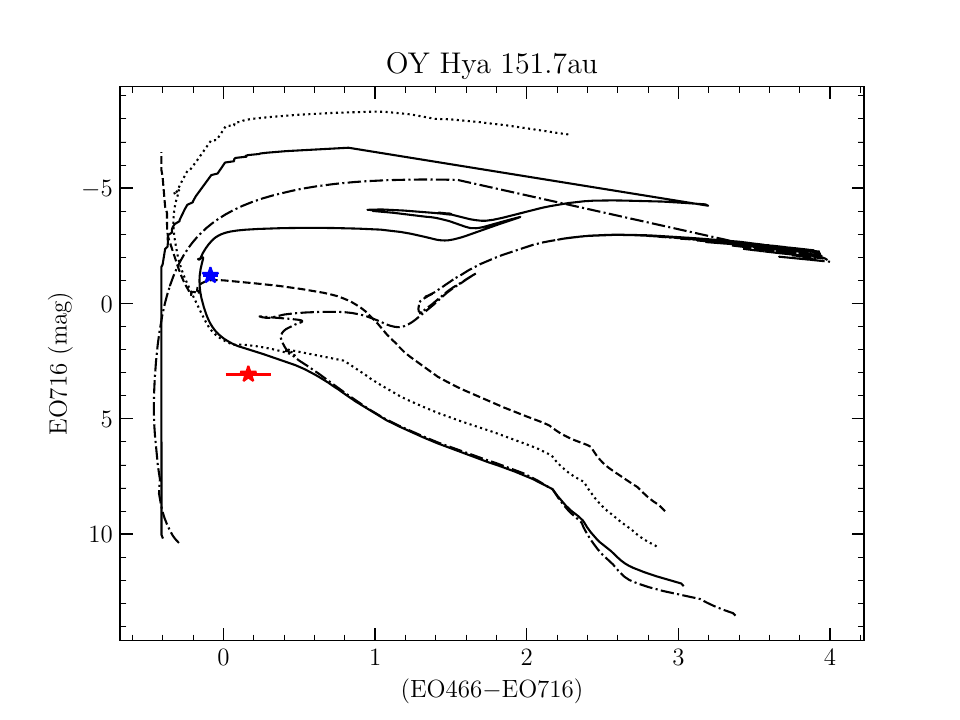}
\includegraphics[width=0.4\textwidth]{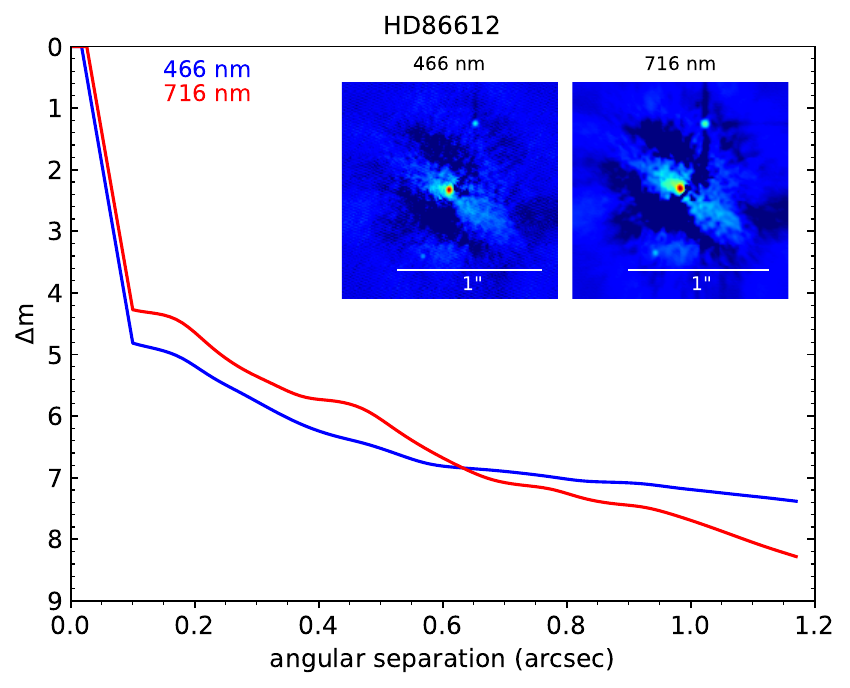}
\includegraphics[width=0.45\textwidth]{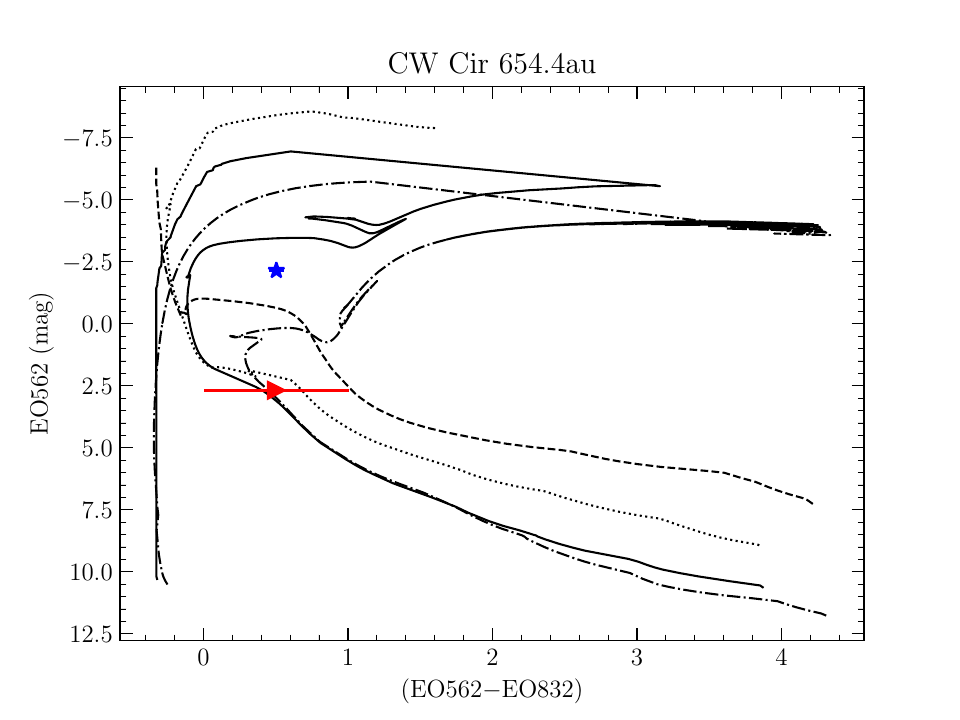}
\includegraphics[width=0.4\textwidth]{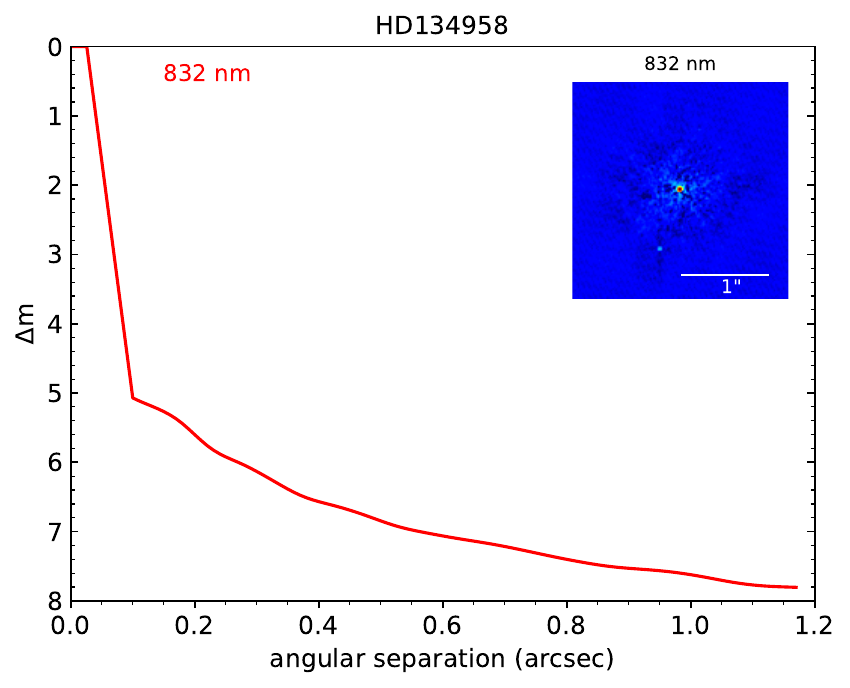}
\includegraphics[width=0.45\textwidth]{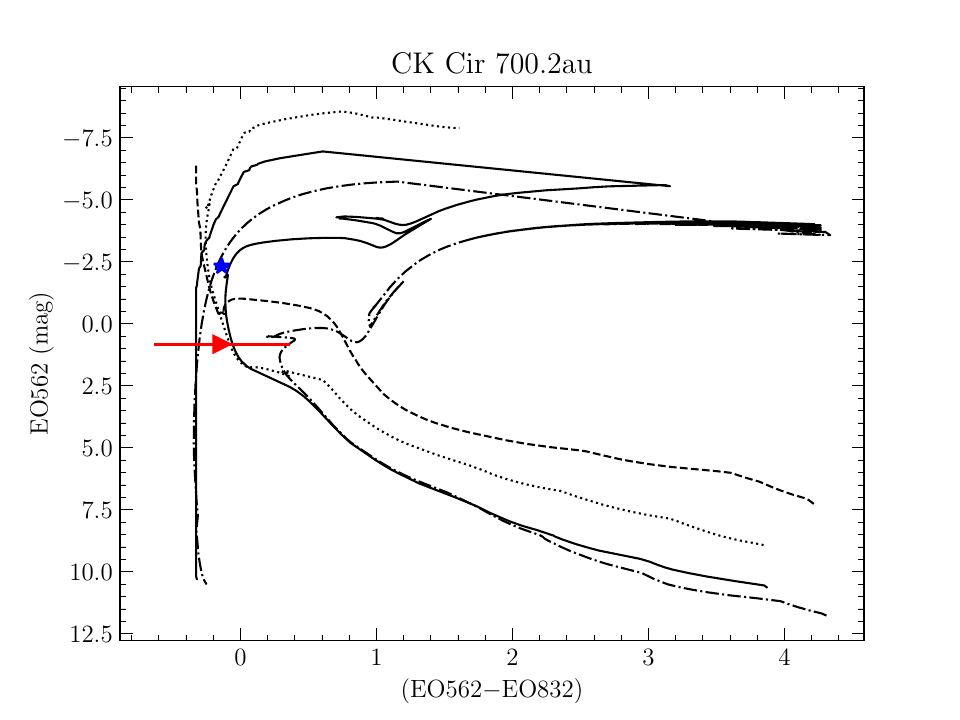}
\includegraphics[width=0.4\textwidth]{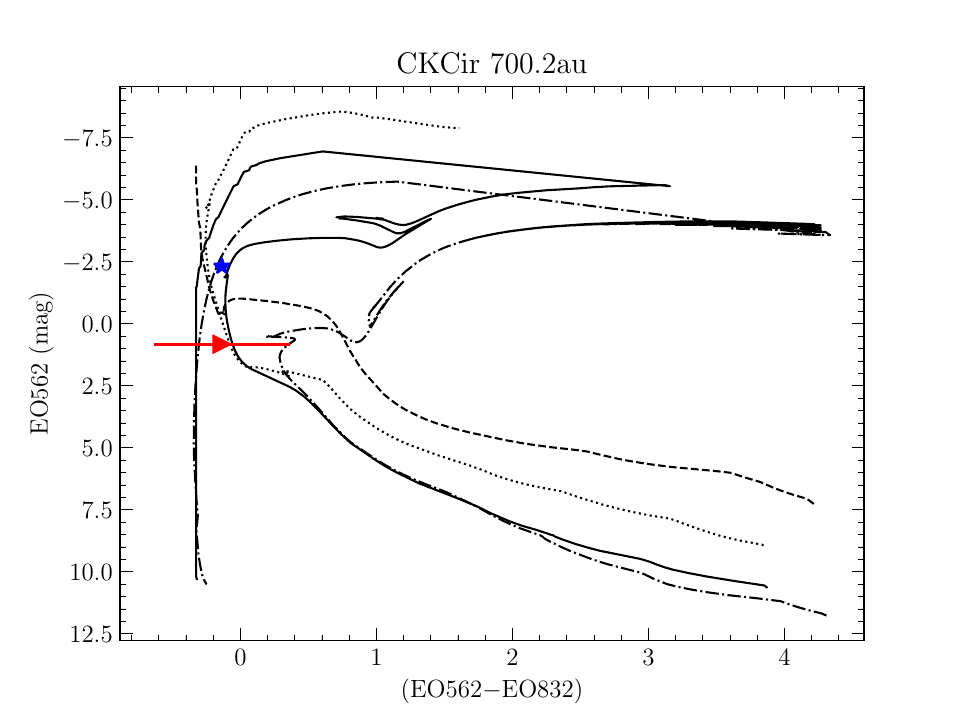}
\caption{Color-magnitude diagrams (left panel) and reconstructed images and contrast curves (right panel) of multiples detected in two speckle filters. \cite{mesa} isochrones computed in the appropriate filters are plotted for 1, 10, 100 Myrs using dashed, dotted, and solid lines. The dashed dotted line is for 1 Gyr. Stars with magnitude in only one filter are assumed to have the same $\Delta m$ in the other for plotting purposes, and are marked by a caret with a color error of 0.2\,mag. }
\label{cmds}
\end{figure*}

\begin{figure*}
\centering
\includegraphics[width=0.5\textwidth]{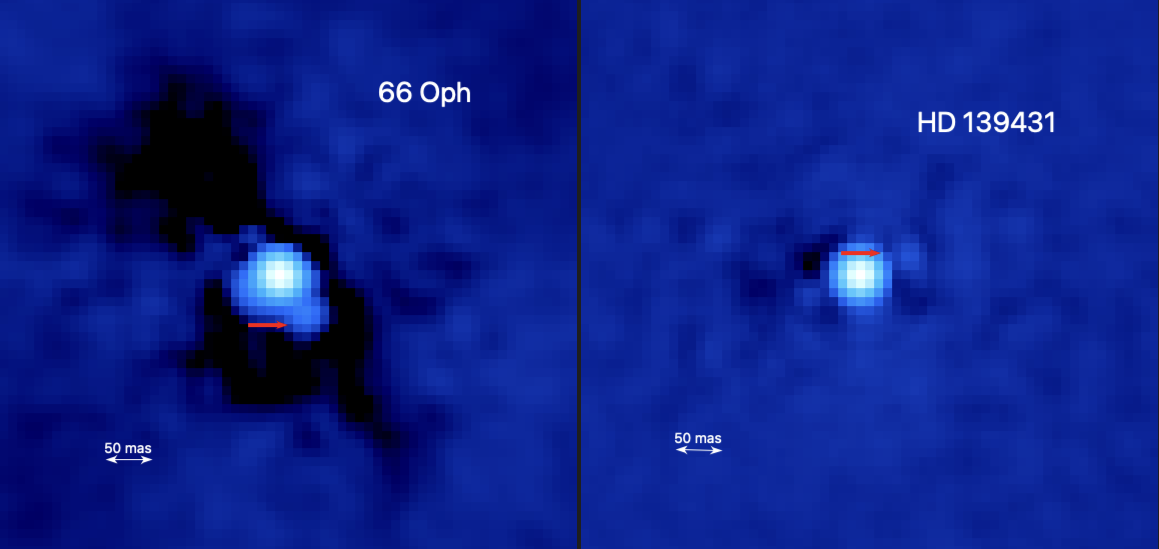}
\includegraphics[width=0.5\textwidth]{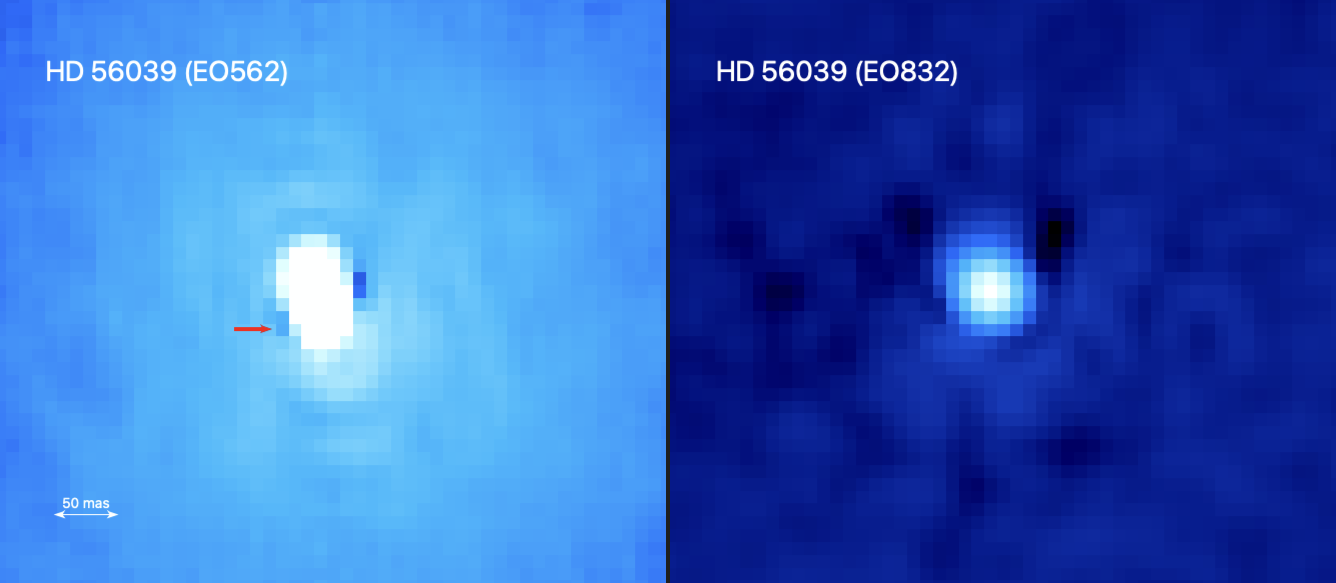}
\caption{Zoomed-in reconstructed images for very close companions. The companion location is shown by the red arrow for 66 Oph (top left), HD\,139431 (top right) in the EO832\,filter, and in both EO\,562 and EO\,832 for HD\,56039 in the bottom panel.}
\label{reconstim}
\end{figure*}

\noindent
{\bf 66\,Oph} (HD\,164284, HIP\,88149, HR\,6712, WDS\,18003\,+0422) is a known binary star previously described in \cite{horch, hutter}, and first identified in \cite{oudmaijer}. Our measurements taken on 26 June 2021 with $`$Alopeke agree within errors with the orbital parameters determined by \cite{hutter}. The secondary found by \cite{hutter} was suggested to be a main-sequence B8 spectral type. The period ($\sim$60\,yrs) is sufficiently large to prevent interaction, although the closest approach of the two stars is around 10\,au. Based on the $\Delta m$ from two speckle filters, we compute the position of the secondary on the color-magnitude diagram (CMD). We adopt the solar metallicity MESA (Modules for Experiments in Stellar Astrophysics)  Isochrones \& Stellar Tracks (MIST) stellar tracks and isochrones \citep{mesa} and the {\it Gaia} DR3 extinction \cite{gaianss} of $A_0$=0.186\,mag to estimate the mass and age of the binary components (see Panel(a) of Fig.\,\ref{cmds}). Since speckle photometry is not absolutely calibrated, we use the {\it Gaia} DR3 spectrum for absolute flux calibration of the primary. 
Our results suggest a mass of 8.4 and 3.5\,$M_{\odot}$ for the primary and secondary, respectively, assuming a coeval age of 10\,Myr to agree with the \cite{hutter} spectral classification of B2 primary and B7/B8 secondary. The companion is visualized in the zoomed-in reconstructed images given in Fig.\,\ref{reconstim} for 66\,Oph, and other companions closer than 0.05$\arcsec$.


It has been recently shown that Be stars are on average brighter than their B-type spectral counterparts in broad $G$ or $V$-band photometry by $\sim$0.5\,mag \citep{issac}, but are similar in color. 
If the $\Delta m$ of the companions are systematically over-estimated by this amount, this would effect mainly the magnitude and not the color of our sources, and lead to an over-estimate of spectral type by one spectral subclass for companions \citep{pecaut}.

\begin{figure*}
\centering
\includegraphics[width=1\textwidth]{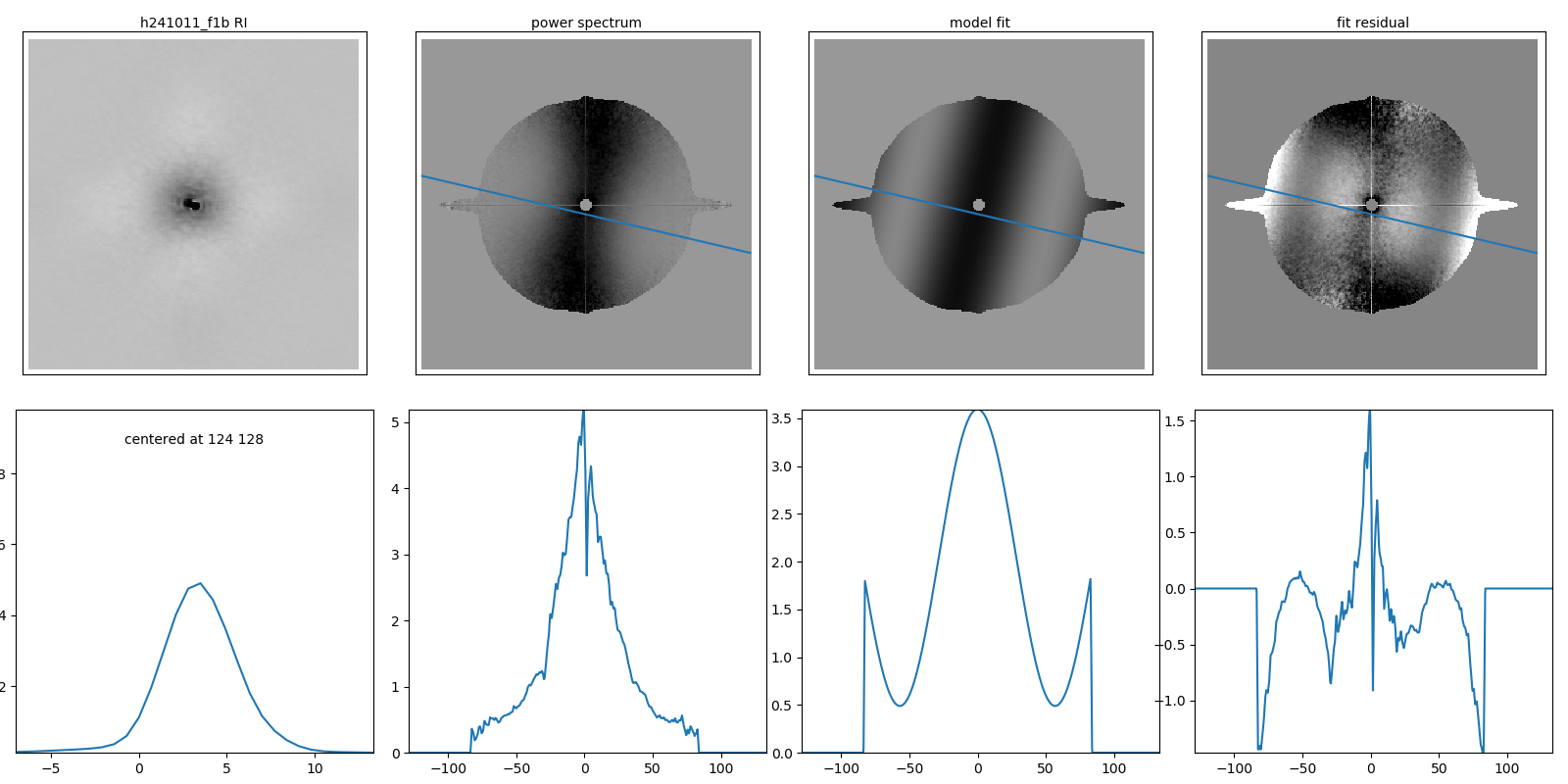}
\caption{The image, power spectrum, model fit (accounting for the point source standard) and the residual are shown from the reduction steps for HD\,56039 in the blue filter (top). The bottom panel shows the slice across the image (along the blue line in the top panel). The power spectrum fringe indicates the close-in companion.}
\label{powerspec}
\end{figure*}

\vspace{10pt}
\noindent
{\bf HD\,56039} (HIP\,35059) was detected as multiple in speckle imaging. The companion to the primary is 11.8\,au away assuming the Gaia\,DR3 distance, and has not been previously reported on in the literature. The secondary is detected only at 562\,nm, with a small magnitude difference ($\Delta m$=\,0.84\,mag), at very close separation of 21\,mas. It is likely that the secondary is beyond the detection limit at 832\,nm given its closeness, and based on the detection in the bluer filter likely a main sequence star close to the primary. The power spectrum in both channels, produced as part of the data reduction pipeline is shown in Fig.\,\ref{powerspec}. A blue only detection could also be because of other reasons, such as maybe a single bright blue emission line, or not a Planck spectral energy distribution to be detected. The object should be re-observed to confirm this detection.  
Based on the stellar mass, and assuming zero eccentricity, the secondary must have a period of at least 6000 days. Assuming the magnitude limit for the secondary magnitude at 832\,nm, it is most likely a B6 star on the main sequence, located too close to the primary. Further observation of this target are essential to confirm the orbital period, and the companion's properties.

\vspace{10pt}
\noindent
{\bf HD\,139431} is an early type Be star with a previously unknown companion. Here, we detect a companion within 37.5\,au in only EO\,832 (observed only in this filter), allowing us to place limits on the secondary. Note the HD\,139431 has different spectral classification in the UV (B5Ve) compared to B2Ve/B3Ve from the optical \citep{skiff}, indicating potentially the presence of an evolved hot companion. Interpolating against the \cite{mesa} stellar models and tracks, we compute the primary mass to be 5.6\,$M_{\odot}$, with a mass ratio around $q$=0.4, suggesting an early A secondary if on the main sequence. For stars with no magnitudes in one filter, we assume the color difference to be zero to place them on the color-magnitude diagram. 

\vspace{10pt}
\noindent
{\bf FV\,CMa} (HD\,54309, HIP\,34360, HR\,2690, WDS\,J07074-2350) is a previously detected binary, with at least 8 epochs reported in the literature (see \citealt{toko21} for a summary). However, no orbital elements, or nature of the secondary are constrained in the literature. The currently available literature data are insufficient to compute reliable orbital periods. Adopting the same methods as previously, we constrain the secondary to a B8V spectral type assuming a coeval age to the primary (10\,Myr).

\vspace{10pt}
\noindent
{\bf QR\,Vul} (HD192685, WDS J20153$+$2536, HIP\,99824) is a previously known binary in \cite{hartkopf}. It also matches closely in separation with the observations using the PISCO (Pupil Interferometry Speckle COronagraph) speckle images \citep[see summary in][]{scardia}, who find the companion at multiple epochs (6 in total, when including this work). This companion has been known since 1879 and has $\sim$30 observations noted in the WDS catalog.

Based on the available data and placing the objects on the CMD, the primary is 4.7\,$M_{\odot}$ star (spectral type of B3), with the secondary having a mass of 1.8\,$M_{\odot}$ (A3 type), assuming a coeval age of 100\,Myr.

\vspace{10pt}
\noindent
{\bf V2120\,Cyg} (HD\,194883, HIP\,100744) is a newly identified companion with no literature detections. The companion is located 75\,au away, and appears to be on the main sequence as well. Assuming a coeval age, we suggest a spectral type of B6V for the companion based on its mass.  

\vspace{10pt}
\noindent
{\bf CU\,Cir} (HD\,133495, HIP\,74011) is newly identified companion, observed and detected only in the red camera. The companion is similar in mass (assuming coeval ages), given the small $\Delta m$ of 0.68\,mag, and in close orbit (86\,au), but not interacting.

\vspace{10pt}
\noindent
{\bf HD\,59498} (HIP\,36397) is newly identified companion detected in the EO\,832 filter only, but observed with both cameras. The companion is faint, and must be at least mid-G given the $\Delta m$ (estimated mass $\sim$0.8\,$M_{\odot}$).

\vspace{10pt}
\noindent
{\bf OY\,Hya} (HD\,86612, HIP\,48943, HR\,3946) has been previously detected as a binary in the speckle observations of \cite{toko21}, with two epochs detected previously, including in \cite{oudmaijer}. The companion was first resolved in Hipparcos data (1991.25). The companion is around 150\,au away, and is very faint. It is not bluer than the primary, however is sufficiently blue that it is either likely a faint sdB star (unlikely given its distance to the primary), or a late type G-type ($\sim$1\,$M_{\odot}$) main-sequence companion. 

\vspace{10pt}
\noindent
{\bf CW\,Cir} (HD\,134958, HIP\,74654) has a companion located 650\,au away, observed and identified only in EO\,832. It is included in the catalog of \cite{bod20} but they do not detect the binary given it's distance from the primary, and their focus on close spectroscopic companions. 
The companion is faint compared to the binary, but must be an early A spectral type based on the difference in magnitudes.

\vspace{10pt}
\noindent
{\bf CK\,Cir} (HD\,128293, HIP\,71668) is a newly identified companion found only in the red camera (observed only with). The companion is between late B-early A, and is located 700\,au away.

\vspace{10pt}
\noindent
{\bf QV Tel} (HD\,167128, HR\,6819). Although not part of our final catalog of binaries, we observed QV\,Tel as part of our observations and detected no binarity in speckle imaging in only the EO\,832 filter, agreeing with recent suggestion of a inner stripped star \citep{frost22}. We suggest that the detection in \cite{klement20} could be of the reference star, HR\,6622 which has a newly detected companion at similar separation and PA as reported for HR\,6819, and is now a quadruple (prv. comm. A. Tokovinin) as indicated by a re-analysis of the existing multiple epoch data. Further deeper data of the primary in multiple filters can help ascertain this.

\bibliography{sample631}{}

\begin{thebibliography}{}
\expandafter\ifx\csname natexlab\endcsname\relax\def\natexlab#1{#1}\fi
\providecommand{\url}[1]{\href{#1}{#1}}
\providecommand{\dodoi}[1]{doi:~\href{http://doi.org/#1}{\nolinkurl{#1}}}
\providecommand{\doeprint}[1]{\href{http://ascl.net/#1}{\nolinkurl{http://ascl.net/#1}}}
\providecommand{\doarXiv}[1]{\href{https://arxiv.org/abs/#1}{\nolinkurl{https://arxiv.org/abs/#1}}}

\bibitem[{{Abt}(2005)}]{abt05}
{Abt}, H.~A. 2005, \apj, 629, 507, \dodoi{10.1086/431207}

\bibitem[{{Abt} \& {Cardona}(1984)}]{abt84}
{Abt}, H.~A., \& {Cardona}, O. 1984, \apj, 285, 190, \dodoi{10.1086/162490}

\bibitem[{{Abt} \& {Levy}(1978)}]{abtspec}
{Abt}, H.~A., \& {Levy}, S.~G. 1978, \apjs, 36, 241, \dodoi{10.1086/190498}

\bibitem[{{Bailer-Jones} {et~al.}(2021){Bailer-Jones}, {Rybizki}, {Fouesneau}, {Demleitner}, \& {Andrae}}]{bailerjones21}
{Bailer-Jones}, C.~A.~L., {Rybizki}, J., {Fouesneau}, M., {Demleitner}, M., \& {Andrae}, R. 2021, \aj, 161, 147, \dodoi{10.3847/1538-3881/abd806}

\bibitem[{{Belokurov} {et~al.}(2020){Belokurov}, {Penoyre}, {Oh}, {Iorio}, {Hodgkin}, {Evans}, {Everall}, {Koposov}, {Tout}, {Izzard}, {Clarke}, \& {Brown}}]{berklov}
{Belokurov}, V., {Penoyre}, Z., {Oh}, S., {et~al.} 2020, \mnras, 496, 1922, \dodoi{10.1093/mnras/staa1522}

\bibitem[{{Berger} \& {Gies}(2001)}]{berger}
{Berger}, D.~H., \& {Gies}, D.~R. 2001, \apj, 555, 364, \dodoi{10.1086/321461}

\bibitem[{{Bjorkman} {et~al.}(2002){Bjorkman}, {Miroshnichenko}, {McDavid}, \& {Pogrosheva}}]{bjorkman}
{Bjorkman}, K.~S., {Miroshnichenko}, A.~S., {McDavid}, D., \& {Pogrosheva}, T.~M. 2002, \apj, 573, 812, \dodoi{10.1086/340751}

\bibitem[{{Bodensteiner} {et~al.}(2020{\natexlab{a}}){Bodensteiner}, {Shenar}, \& {Sana}}]{bod20}
{Bodensteiner}, J., {Shenar}, T., \& {Sana}, H. 2020{\natexlab{a}}, \aap, 641, A42, \dodoi{10.1051/0004-6361/202037640}

\bibitem[{{Bodensteiner} {et~al.}(2020{\natexlab{b}}){Bodensteiner}, {Shenar}, {Mahy}, {Fabry}, {Marchant}, {Abdul-Masih}, {Banyard}, {Bowman}, {Dsilva}, {Frost}, {Hawcroft}, {Reggiani}, \& {Sana}}]{bodqvtel}
{Bodensteiner}, J., {Shenar}, T., {Mahy}, L., {et~al.} 2020{\natexlab{b}}, \aap, 641, A43, \dodoi{10.1051/0004-6361/202038682}

\bibitem[{{Boubert} \& {Evans}(2018)}]{boubert}
{Boubert}, D., \& {Evans}, N.~W. 2018, \mnras, 477, 5261, \dodoi{10.1093/mnras/sty980}

\bibitem[{{Chen} {et~al.}(2016){Chen}, {Liu}, \& {Shan}}]{chen16}
{Chen}, P.~S., {Liu}, J.~Y., \& {Shan}, H.~G. 2016, \mnras, 463, 1162, \dodoi{10.1093/mnras/stw1757}

\bibitem[{{Cifuentes} {et~al.}(2025){Cifuentes}, {Caballero}, {Gonz{\'a}lez-Payo}, {Amado}, {B{\'e}jar}, {Burgasser}, {Cort{\'e}s-Contreras}, {Lodieu}, {Montes}, {Quirrenbach}, {Reiners}, {Ribas}, {Sanz-Forcada}, {Seifert}, \& {Zapatero Osorio}}]{cifuentes}
{Cifuentes}, C., {Caballero}, J.~A., {Gonz{\'a}lez-Payo}, J., {et~al.} 2025, \aap, 693, A228, \dodoi{10.1051/0004-6361/202452527}

\bibitem[{{Correia} {et~al.}(2006){Correia}, {Zinnecker}, {Ratzka}, \& {Sterzik}}]{2006A&A...459..909C}
{Correia}, S., {Zinnecker}, H., {Ratzka}, T., \& {Sterzik}, M.~F. 2006, \aap, 459, 909, \dodoi{10.1051/0004-6361:20065545}

\bibitem[{{Cruz-Gonz{\'a}lez} {et~al.}(1974){Cruz-Gonz{\'a}lez}, {Recillas-Cruz}, {Costero}, {Peimbert}, \& {Torres-Peimbert}}]{1974RMxAA...1..211C}
{Cruz-Gonz{\'a}lez}, C., {Recillas-Cruz}, E., {Costero}, R., {Peimbert}, M., \& {Torres-Peimbert}, S. 1974, \rmxaa, 1, 211

\bibitem[{{Dodd} {et~al.}(2024){Dodd}, {Oudmaijer}, {Radley}, {Vioque}, \& {Frost}}]{dodd}
{Dodd}, J.~M., {Oudmaijer}, R.~D., {Radley}, I.~C., {Vioque}, M., \& {Frost}, A.~J. 2024, \mnras, 527, 3076, \dodoi{10.1093/mnras/stad3105}

\bibitem[{{Dotter}(2016)}]{mesa}
{Dotter}, A. 2016, \apjs, 222, 8, \dodoi{10.3847/0067-0049/222/1/8}

\bibitem[{{Evans} \& {Edwards}(1981)}]{evans1981}
{Evans}, D.~S., \& {Edwards}, D.~A. 1981, \aj, 86, 1277, \dodoi{10.1086/113008}

\bibitem[{{Fitton} {et~al.}(2022){Fitton}, {Tofflemire}, \& {Kraus}}]{fitton}
{Fitton}, S., {Tofflemire}, B.~M., \& {Kraus}, A.~L. 2022, Research Notes of the American Astronomical Society, 6, 18, \dodoi{10.3847/2515-5172/ac4bb7}

\bibitem[{{Frost} {et~al.}(2022){Frost}, {Bodensteiner}, {Rivinius}, {Baade}, {Merand}, {Selman}, {Abdul-Masih}, {Banyard}, {Bordier}, {Dsilva}, {Hawcroft}, {Mahy}, {Reggiani}, {Shenar}, {Cabezas}, {Hadrava}, {Heida}, {Klement}, \& {Sana}}]{frost22}
{Frost}, A.~J., {Bodensteiner}, J., {Rivinius}, T., {et~al.} 2022, \aap, 659, L3, \dodoi{10.1051/0004-6361/202143004}

\bibitem[{{Frost} {et~al.}(2025){Frost}, {Sana}, {Le Bouquin}, {Perets}, {Bodensteiner}, {Igoshev}, {Banyard}, {Mahy}, {M{\'e}rand}, \& {Ram{\'\i}rez-Agudelo}}]{frost25}
{Frost}, A.~J., {Sana}, H., {Le Bouquin}, J.-B., {et~al.} 2025, arXiv e-prints, arXiv:2505.02300, \dodoi{10.48550/arXiv.2505.02300}

\bibitem[{{Gaia Collaboration} {et~al.}(2018){Gaia Collaboration}, {Brown}, {Vallenari}, {Prusti}, {de Bruijne}, {Babusiaux}, {Bailer-Jones}, {Biermann}, {Evans}, {Eyer}, {Jansen}, {Jordi}, {Klioner}, {Lammers}, {Lindegren}, {Luri}, {Mignard}, {Panem}, {Pourbaix}, {Randich}, {Sartoretti}, {Siddiqui}, {Soubiran}, {van Leeuwen}, {Walton}, {Arenou}, {Bastian}, {Cropper}, {Drimmel}, {Katz}, {Lattanzi}, {Bakker}, {Cacciari}, {Casta{\~n}eda}, {Chaoul}, {Cheek}, {De Angeli}, {Fabricius}, {Guerra}, {Holl}, {Masana}, {Messineo}, {Mowlavi}, {Nienartowicz}, {Panuzzo}, {Portell}, {Riello}, {Seabroke}, {Tanga}, {Th{\'e}venin}, {Gracia-Abril}, {Comoretto}, {Garcia-Reinaldos}, {Teyssier}, {Altmann}, {Andrae}, {Audard}, {Bellas-Velidis}, {Benson}, {Berthier}, {Blomme}, {Burgess}, {Busso}, {Carry}, {Cellino}, {Clementini}, {Clotet}, {Creevey}, {Davidson}, {De Ridder}, {Delchambre}, {Dell'Oro}, {Ducourant}, {Fern{\'a}ndez-Hern{\'a}ndez}, {Fouesneau}, {Fr{\'e}mat}, {Galluccio}, {Garc{\'\i}a-Torres},
  {Gonz{\'a}lez-N{\'u}{\~n}ez}, {Gonz{\'a}lez-Vidal}, {Gosset}, {Guy}, {Halbwachs}, {Hambly}, {Harrison}, {Hern{\'a}ndez}, {Hestroffer}, {Hodgkin}, {Hutton}, {Jasniewicz}, {Jean-Antoine-Piccolo}, {Jordan}, {Korn}, {Krone-Martins}, {Lanzafame}, {Lebzelter}, {L{\"o}ffler}, {Manteiga}, {Marrese}, {Mart{\'\i}n-Fleitas}, {Moitinho}, {Mora}, {Muinonen}, {Osinde}, {Pancino}, {Pauwels}, {Petit}, {Recio-Blanco}, {Richards}, {Rimoldini}, {Robin}, {Sarro}, {Siopis}, {Smith}, {Sozzetti}, {S{\"u}veges}, {Torra}, {van Reeven}, {Abbas}, {Abreu Aramburu}, {Accart}, {Aerts}, {Altavilla}, {{\'A}lvarez}, {Alvarez}, {Alves}, {Anderson}, {Andrei}, {Anglada Varela}, {Antiche}, {Antoja}, {Arcay}, {Astraatmadja}, {Bach}, {Baker}, {Balaguer-N{\'u}{\~n}ez}, {Balm}, {Barache}, {Barata}, {Barbato}, {Barblan}, {Barklem}, {Barrado}, {Barros}, {Barstow}, {Bartholom{\'e} Mu{\~n}oz}, {Bassilana}, {Becciani}, {Bellazzini}, {Berihuete}, {Bertone}, {Bianchi}, {Bienaym{\'e}}, {Blanco-Cuaresma}, {Boch}, {Boeche}, {Bombrun}, {Borrachero},
  {Bossini}, {Bouquillon}, {Bourda}, {Bragaglia}, {Bramante}, {Breddels}, {Bressan}, {Brouillet}, {Br{\"u}semeister}, {Brugaletta}, {Bucciarelli}, {Burlacu}, {Busonero}, {Butkevich}, {Buzzi}, {Caffau}, {Cancelliere}, {Cannizzaro}, {Cantat-Gaudin}, {Carballo}, {Carlucci}, {Carrasco}, {Casamiquela}, {Castellani}, {Castro-Ginard}, {Charlot}, {Chemin}, {Chiavassa}, {Cocozza}, {Costigan}, {Cowell}, {Crifo}, {Crosta}, {Crowley}, {Cuypers}, {Dafonte}, {Damerdji}, {Dapergolas}, {David}, {David}, {de Laverny}, \& {De Luise}}]{dr2}
{Gaia Collaboration}, {Brown}, A.~G.~A., {Vallenari}, A., {et~al.} 2018, \aap, 616, A1, \dodoi{10.1051/0004-6361/201833051}

\bibitem[{{Gaia Collaboration} {et~al.}(2023){Gaia Collaboration}, {Vallenari}, {Brown}, {Prusti}, {de Bruijne}, {Arenou}, {Babusiaux}, {Biermann}, {Creevey}, {Ducourant}, {Evans}, {Eyer}, {Guerra}, {Hutton}, {Jordi}, {Klioner}, {Lammers}, {Lindegren}, {Luri}, {Mignard}, {Panem}, {Pourbaix}, {Randich}, {Sartoretti}, {Soubiran}, {Tanga}, {Walton}, {Bailer-Jones}, {Bastian}, {Drimmel}, {Jansen}, {Katz}, {Lattanzi}, {van Leeuwen}, {Bakker}, {Cacciari}, {Casta{\~n}eda}, {De Angeli}, {Fabricius}, {Fouesneau}, {Fr{\'e}mat}, {Galluccio}, {Guerrier}, {Heiter}, {Masana}, {Messineo}, {Mowlavi}, {Nicolas}, {Nienartowicz}, {Pailler}, {Panuzzo}, {Riclet}, {Roux}, {Seabroke}, {Sordo}, {Th{\'e}venin}, {Gracia-Abril}, {Portell}, {Teyssier}, {Altmann}, {Andrae}, {Audard}, {Bellas-Velidis}, {Benson}, {Berthier}, {Blomme}, {Burgess}, {Busonero}, {Busso}, {C{\'a}novas}, {Carry}, {Cellino}, {Cheek}, {Clementini}, {Damerdji}, {Davidson}, {de Teodoro}, {Nu{\~n}ez Campos}, {Delchambre}, {Dell'Oro}, {Esquej},
  {Fern{\'a}ndez-Hern{\'a}ndez}, {Fraile}, {Garabato}, {Garc{\'\i}a-Lario}, {Gosset}, {Haigron}, {Halbwachs}, {Hambly}, {Harrison}, {Hern{\'a}ndez}, {Hestroffer}, {Hodgkin}, {Holl}, {Jan{\ss}en}, {Jevardat de Fombelle}, {Jordan}, {Krone-Martins}, {Lanzafame}, {L{\"o}ffler}, {Marchal}, {Marrese}, {Moitinho}, {Muinonen}, {Osborne}, {Pancino}, {Pauwels}, {Recio-Blanco}, {Reyl{\'e}}, {Riello}, {Rimoldini}, {Roegiers}, {Rybizki}, {Sarro}, {Siopis}, {Smith}, {Sozzetti}, {Utrilla}, {van Leeuwen}, {Abbas}, {{\'A}brah{\'a}m}, {Abreu Aramburu}, {Aerts}, {Aguado}, {Ajaj}, {Aldea-Montero}, {Altavilla}, {{\'A}lvarez}, {Alves}, {Anders}, {Anderson}, {Anglada Varela}, {Antoja}, {Baines}, {Baker}, {Balaguer-N{\'u}{\~n}ez}, {Balbinot}, {Balog}, {Barache}, {Barbato}, {Barros}, {Barstow}, {Bartolom{\'e}}, {Bassilana}, {Bauchet}, {Becciani}, {Bellazzini}, {Berihuete}, {Bernet}, {Bertone}, {Bianchi}, {Binnenfeld}, {Blanco-Cuaresma}, {Blazere}, {Boch}, {Bombrun}, {Bossini}, {Bouquillon}, {Bragaglia}, {Bramante}, {Breedt},
  {Bressan}, {Brouillet}, {Brugaletta}, {Bucciarelli}, {Burlacu}, {Butkevich}, {Buzzi}, {Caffau}, {Cancelliere}, {Cantat-Gaudin}, {Carballo}, {Carlucci}, {Carnerero}, {Carrasco}, {Casamiquela}, {Castellani}, {Castro-Ginard}, {Chaoul}, {Charlot}, {Chemin}, {Chiaramida}, {Chiavassa}, {Chornay}, {Comoretto}, {Contursi}, {Cooper}, {Cornez}, {Cowell}, {Crifo}, {Cropper}, {Crosta}, {Crowley}, {Dafonte}, {Dapergolas}, {David}, {David}, {de Laverny}, {De Luise}, \& {De March}}]{gaianss}
{Gaia Collaboration}, {Vallenari}, A., {Brown}, A.~G.~A., {et~al.} 2023, \aap, 674, A1, \dodoi{10.1051/0004-6361/202243940}

\bibitem[{{Guerrero} {et~al.}(2025){Guerrero}, {Souza}, {Borges Fernandes}, \& {Guajardo Jurado}}]{2025AJ....169..251G}
{Guerrero}, C.~A., {Souza}, T.~B., {Borges Fernandes}, M., \& {Guajardo Jurado}, A.~D. 2025, \aj, 169, 251, \dodoi{10.3847/1538-3881/adbf11}

\bibitem[{{Hartkopf} {et~al.}(2012){Hartkopf}, {Tokovinin}, \& {Mason}}]{2012AJ....143...42H}
{Hartkopf}, W.~I., {Tokovinin}, A., \& {Mason}, B.~D. 2012, \aj, 143, 42, \dodoi{10.1088/0004-6256/143/2/42}

\bibitem[{{Hartkopf} {et~al.}(2000){Hartkopf}, {Mason}, {McAlister}, {Roberts}, {Turner}, {ten Brummelaar}, {Prieto}, {Ling}, \& {Franz}}]{hartkopf}
{Hartkopf}, W.~I., {Mason}, B.~D., {McAlister}, H.~A., {et~al.} 2000, \aj, 119, 3084, \dodoi{10.1086/301402}

\bibitem[{{Horch} {et~al.}(2001){Horch}, {Ninkov}, \& {Franz}}]{horch01}
{Horch}, E., {Ninkov}, Z., \& {Franz}, O.~G. 2001, \aj, 121, 1583, \dodoi{10.1086/319423}

\bibitem[{{Horch} {et~al.}(2011){Horch}, {Gomez}, {Sherry}, {Howell}, {Ciardi}, {Anderson}, \& {van Altena}}]{horch11}
{Horch}, E.~P., {Gomez}, S.~C., {Sherry}, W.~H., {et~al.} 2011, \aj, 141, 45, \dodoi{10.1088/0004-6256/141/2/45}

\bibitem[{{Horch} {et~al.}(2020){Horch}, {van Belle}, {Davidson}, {Willmarth}, {Fekel}, {Muterspaugh}, {Casetti-Dinescu}, {Hahne}, {Granucci}, {Clark}, {Winters}, {Rupert}, {Weiss}, {Colton}, {Nusdeo}, \& {Henry}}]{horch}
{Horch}, E.~P., {van Belle}, G.~T., {Davidson}, Jr., J.~W., {et~al.} 2020, \aj, 159, 233, \dodoi{10.3847/1538-3881/ab87a6}

\bibitem[{{Howell} {et~al.}(2011){Howell}, {Everett}, {Sherry}, {Horch}, \& {Ciardi}}]{howell11}
{Howell}, S.~B., {Everett}, M.~E., {Sherry}, W., {Horch}, E., \& {Ciardi}, D.~R. 2011, \aj, 142, 19, \dodoi{10.1088/0004-6256/142/1/19}

\bibitem[{{Howell} {et~al.}(2025){Howell}, {Mart{\'\i}nez-V{\'a}zquez}, {Furlan}, {Scott}, {Matson}, {Littlefield}, {Clark}, {Lester}, {Hartman}, {Ciardi}, \& {Deveny}}]{howell25}
{Howell}, S.~B., {Mart{\'\i}nez-V{\'a}zquez}, C.~E., {Furlan}, E., {et~al.} 2025, Frontiers in Astronomy and Space Sciences, 12, 1608411, \dodoi{10.3389/fspas.2025.1608411}

\bibitem[{{Hutter} {et~al.}(2021){Hutter}, {Tycner}, {Zavala}, {Benson}, {Hummel}, \& {Zirm}}]{hutter}
{Hutter}, D.~J., {Tycner}, C., {Zavala}, R.~T., {et~al.} 2021, \apjs, 257, 69, \dodoi{10.3847/1538-4365/ac23cb}

\bibitem[{{Kalari}(2019)}]{ngc6383}
{Kalari}, V.~M. 2019, \mnras, 484, 5102, \dodoi{10.1093/mnras/stz250}

\bibitem[{{Kalari} {et~al.}(2024){Kalari}, {Salinas}, {Zinnecker}, {Rubio}, {Herczeg}, \& {Andersen}}]{kalari24}
{Kalari}, V.~M., {Salinas}, R., {Zinnecker}, H., {et~al.} 2024, \apj, 972, 3, \dodoi{10.3847/1538-4357/ad5bd9}

\bibitem[{{Kalari} {et~al.}(2019){Kalari}, {Vink}, {de Wit}, {Bastian}, \& {M{\'e}ndez}}]{kalarirun}
{Kalari}, V.~M., {Vink}, J.~S., {de Wit}, W.~J., {Bastian}, N.~J., \& {M{\'e}ndez}, R.~A. 2019, \aap, 625, L2, \dodoi{10.1051/0004-6361/201935107}

\bibitem[{{Kervella} {et~al.}(2021){Kervella}, {Arenou}, \& {Thevenin}}]{kervella}
{Kervella}, P., {Arenou}, F., \& {Thevenin}, F. 2021, {VizieR Online Data Catalog: Stellar and substellar companions from Gaia EDR3 (Kervella+, 2022)}, VizieR On-line Data Catalog: J/A+A/657/A7. Originally published in: 2022A\&A...657A...7K, \dodoi{10.26093/cds/vizier.36570007}

\bibitem[{{Klement} {et~al.}(2021{\natexlab{a}}){Klement}, {Scott}, {Rivinius}, {Baade}, \& {Hadrava}}]{klement20}
{Klement}, R., {Scott}, N., {Rivinius}, T., {Baade}, D., \& {Hadrava}, P. 2021{\natexlab{a}}, The Astronomer's Telegram, 14340, 1

\bibitem[{{Klement} {et~al.}(2021{\natexlab{b}}){Klement}, {Hadrava}, {Rivinius}, {Baade}, {Cabezas}, {Heida}, {Schaefer}, {Gardner}, {Gies}, {Anugu}, {Lanthermann}, {Davies}, {Anderson}, {Monnier}, {Ennis}, {Labdon}, {Setterholm}, {Kraus}, {ten Brummelaar}, \& {Le Bouquin}}]{klement21}
{Klement}, R., {Hadrava}, P., {Rivinius}, T., {et~al.} 2021{\natexlab{b}}, \apj, 916, 24, \dodoi{10.3847/1538-4357/ac062c}

\bibitem[{{Klement} {et~al.}(2022){Klement}, {Schaefer}, {Gies}, {Wang}, {Baade}, {Rivinius}, {Gallenne}, {Carciofi}, {Monnier}, {M{\'e}rand}, {Anugu}, {Kraus}, {Davies}, {Lanthermann}, {Gardner}, {Wysocki}, {Ennis}, {Labdon}, {Setterholm}, \& {Le Bouquin}}]{klement22}
{Klement}, R., {Schaefer}, G.~H., {Gies}, D.~R., {et~al.} 2022, \apj, 926, 213, \dodoi{10.3847/1538-4357/ac4266}

\bibitem[{{Klement} {et~al.}(2024){Klement}, {Rivinius}, {Gies}, {Baade}, {M{\'e}rand}, {Monnier}, {Schaefer}, {Lanthermann}, {Anugu}, {Kraus}, \& {Gardner}}]{klementchara}
{Klement}, R., {Rivinius}, T., {Gies}, D.~R., {et~al.} 2024, \apj, 962, 70, \dodoi{10.3847/1538-4357/ad13ec}

\bibitem[{{Koubsk{\'y}} {et~al.}(2012){Koubsk{\'y}}, {Kotkov{\'a}}, {Votruba}, {{\v{S}}lechta}, \& {Dvo{\v{r}}{\'a}kov{\'a}}}]{koub12}
{Koubsk{\'y}}, P., {Kotkov{\'a}}, L., {Votruba}, V., {{\v{S}}lechta}, M., \& {Dvo{\v{r}}{\'a}kov{\'a}}, {\v{S}}. 2012, \aap, 545, A121, \dodoi{10.1051/0004-6361/201219679}

\bibitem[{{Koubsk{\'y}} {et~al.}(2000){Koubsk{\'y}}, {Harmanec}, {Hubert}, {Floquet}, {Kub{\'a}t}, {Ballereau}, {Chauville}, {Bo{\v{z}}i{\'c}}, {Holmgren}, {Yang}, {Cao}, {Eenens}, {Huang}, \& {Percy}}]{koub00}
{Koubsk{\'y}}, P., {Harmanec}, P., {Hubert}, A.~M., {et~al.} 2000, \aap, 356, 913

\bibitem[{{Lester} {et~al.}(2021){Lester}, {Matson}, {Howell}, {Furlan}, {Gnilka}, {Scott}, {Ciardi}, {Everett}, {Hartman}, \& {Hirsch}}]{2021AJ....162...75L}
{Lester}, K.~V., {Matson}, R.~A., {Howell}, S.~B., {et~al.} 2021, \aj, 162, 75, \dodoi{10.3847/1538-3881/ac0d06}

\bibitem[{{Lindegren} {et~al.}(2018){Lindegren}, {Hern{\'a}ndez}, {Bombrun}, {Klioner}, {Bastian}, {Ramos-Lerate}, {de Torres}, {Steidelm{\"u}ller}, {Stephenson}, {Hobbs}, {Lammers}, {Biermann}, {Geyer}, {Hilger}, {Michalik}, {Stampa}, {McMillan}, {Casta{\~n}eda}, {Clotet}, {Comoretto}, {Davidson}, {Fabricius}, {Gracia}, {Hambly}, {Hutton}, {Mora}, {Portell}, {van Leeuwen}, {Abbas}, {Abreu}, {Altmann}, {Andrei}, {Anglada}, {Balaguer-N{\'u}{\~n}ez}, {Barache}, {Becciani}, {Bertone}, {Bianchi}, {Bouquillon}, {Bourda}, {Br{\"u}semeister}, {Bucciarelli}, {Busonero}, {Buzzi}, {Cancelliere}, {Carlucci}, {Charlot}, {Cheek}, {Crosta}, {Crowley}, {de Bruijne}, {de Felice}, {Drimmel}, {Esquej}, {Fienga}, {Fraile}, {Gai}, {Garralda}, {Gonz{\'a}lez-Vidal}, {Guerra}, {Hauser}, {Hofmann}, {Holl}, {Jordan}, {Lattanzi}, {Lenhardt}, {Liao}, {Licata}, {Lister}, {L{\"o}ffler}, {Marchant}, {Martin-Fleitas}, {Messineo}, {Mignard}, {Morbidelli}, {Poggio}, {Riva}, {Rowell}, {Salguero}, {Sarasso}, {Sciacca}, {Siddiqui}, {Smart},
  {Spagna}, {Steele}, {Taris}, {Torra}, {van Elteren}, {van Reeven}, \& {Vecchiato}}]{lind}
{Lindegren}, L., {Hern{\'a}ndez}, J., {Bombrun}, A., {et~al.} 2018, \aap, 616, A2, \dodoi{10.1051/0004-6361/201832727}

\bibitem[{{Lindegren} {et~al.}(2021){Lindegren}, {Klioner}, {Hern{\'a}ndez}, {Bombrun}, {Ramos-Lerate}, {Steidelm{\"u}ller}, {Bastian}, {Biermann}, {de Torres}, {Gerlach}, {Geyer}, {Hilger}, {Hobbs}, {Lammers}, {McMillan}, {Stephenson}, {Casta{\~n}eda}, {Davidson}, {Fabricius}, {Gracia-Abril}, {Portell}, {Rowell}, {Teyssier}, {Torra}, {Bartolom{\'e}}, {Clotet}, {Garralda}, {Gonz{\'a}lez-Vidal}, {Torra}, {Abbas}, {Altmann}, {Anglada Varela}, {Balaguer-N{\'u}{\~n}ez}, {Balog}, {Barache}, {Becciani}, {Bernet}, {Bertone}, {Bianchi}, {Bouquillon}, {Brown}, {Bucciarelli}, {Busonero}, {Butkevich}, {Buzzi}, {Cancelliere}, {Carlucci}, {Charlot}, {Cioni}, {Crosta}, {Crowley}, {del Peloso}, {del Pozo}, {Drimmel}, {Esquej}, {Fienga}, {Fraile}, {Gai}, {Garcia-Reinaldos}, {Guerra}, {Hambly}, {Hauser}, {Jan{\ss}en}, {Jordan}, {Kostrzewa-Rutkowska}, {Lattanzi}, {Liao}, {Licata}, {Lister}, {L{\"o}ffler}, {Marchant}, {Masip}, {Mignard}, {Mints}, {Molina}, {Mora}, {Morbidelli}, {Murphy}, {Pagani}, {Panuzzo}, {Pe{\~n}alosa
  Esteller}, {Poggio}, {Re Fiorentin}, {Riva}, {Sagrist{\`a} Sell{\'e}s}, {Sanchez Gimenez}, {Sarasso}, {Sciacca}, {Siddiqui}, {Smart}, {Souami}, {Spagna}, {Steele}, {Taris}, {Utrilla}, {van Reeven}, \& {Vecchiato}}]{gaialind}
{Lindegren}, L., {Klioner}, S.~A., {Hern{\'a}ndez}, J., {et~al.} 2021, \aap, 649, A2, \dodoi{10.1051/0004-6361/202039709}

\bibitem[{{Lindroos}(1986)}]{lindroos}
{Lindroos}, K.~P. 1986, \aap, 156, 223

\bibitem[{{Martin} \& {Franchini}(2019)}]{bekldvout}
{Martin}, R.~G., \& {Franchini}, A. 2019, \mnras, 489, 1797, \dodoi{10.1093/mnras/stz2250}

\bibitem[{{Mason} {et~al.}(1997){Mason}, {ten Brummelaar}, {Gies}, {Hartkopf}, \& {Thaller}}]{mason97}
{Mason}, B.~D., {ten Brummelaar}, T., {Gies}, D.~R., {Hartkopf}, W.~I., \& {Thaller}, M.~L. 1997, \aj, 114, 2112, \dodoi{10.1086/118630}

\bibitem[{{Mason} {et~al.}(2022){Mason}, {Wycoff}, {Hartkopf}, {Douglass}, \& {Worley}}]{wds}
{Mason}, B.~D., {Wycoff}, G.~L., {Hartkopf}, W.~I., {Douglass}, G.~G., \& {Worley}, C.~E. 2022, {VizieR Online Data Catalog: The Washington Visual Double Star Catalog (Mason+ 2001-2020)}, VizieR On-line Data Catalog: B/wds. Originally published in: 2001AJ....122.3466M

\bibitem[{{Moe} \& {Di Stefano}(2017)}]{distefano}
{Moe}, M., \& {Di Stefano}, R. 2017, \apjs, 230, 15, \dodoi{10.3847/1538-4365/aa6fb6}

\bibitem[{{Moffat} {et~al.}(1998){Moffat}, {Marchenko}, {Seggewiss}, {van der Hucht}, {Schrijver}, {Stenholm}, {Lundstrom}, {Setia Gunawan}, {Sutantyo}, {van den Heuvel}, {de Cuyper}, \& {Gomez}}]{moffat}
{Moffat}, A.~F.~J., {Marchenko}, S.~V., {Seggewiss}, W., {et~al.} 1998, \aap, 331, 949

\bibitem[{{Naoz}(2016)}]{kozai}
{Naoz}, S. 2016, \araa, 54, 441, \dodoi{10.1146/annurev-astro-081915-023315}

\bibitem[{{Offner} {et~al.}(2023){Offner}, {Moe}, {Kratter}, {Sadavoy}, {Jensen}, \& {Tobin}}]{offner}
{Offner}, S.~S.~R., {Moe}, M., {Kratter}, K.~M., {et~al.} 2023, in Astronomical Society of the Pacific Conference Series, Vol. 534, Protostars and Planets VII, ed. S.~{Inutsuka}, Y.~{Aikawa}, T.~{Muto}, K.~{Tomida}, \& M.~{Tamura}, 275, \dodoi{10.48550/arXiv.2203.10066}

\bibitem[{{Oudmaijer} \& {Parr}(2010)}]{oudmaijer}
{Oudmaijer}, R.~D., \& {Parr}, A.~M. 2010, \mnras, 405, 2439, \dodoi{10.1111/j.1365-2966.2010.16609.x}

\bibitem[{{Pecaut} \& {Mamajek}(2013)}]{pecaut}
{Pecaut}, M.~J., \& {Mamajek}, E.~E. 2013, \apjs, 208, 9, \dodoi{10.1088/0067-0049/208/1/9}

\bibitem[{{Pomohaci} {et~al.}(2019){Pomohaci}, {Oudmaijer}, \& {Goodwin}}]{pomohaci}
{Pomohaci}, R., {Oudmaijer}, R.~D., \& {Goodwin}, S.~P. 2019, \mnras, 484, 226, \dodoi{10.1093/mnras/stz014}

\bibitem[{{Pourbaix} {et~al.}(2004){Pourbaix}, {Tokovinin}, {Batten}, {Fekel}, {Hartkopf}, {Levato}, {Morrell}, {Torres}, \& {Udry}}]{sb9}
{Pourbaix}, D., {Tokovinin}, A.~A., {Batten}, A.~H., {et~al.} 2004, \aap, 424, 727, \dodoi{10.1051/0004-6361:20041213}

\bibitem[{{Radley} {et~al.}(2025){Radley}, {Oudmaijer}, {Vioque}, \& {Dodd}}]{issac}
{Radley}, I.~C., {Oudmaijer}, R.~D., {Vioque}, M., \& {Dodd}, J.~M. 2025, \mnras, 539, 1964, \dodoi{10.1093/mnras/staf618}

\bibitem[{{Richards} {et~al.}(2000){Richards}, {Koubsk{\'y}}, {{\v{S}}imon}, {Peters}, {Hirata}, {{\v{S}}koda}, \& {Masuda}}]{cxdra}
{Richards}, M.~T., {Koubsk{\'y}}, P., {{\v{S}}imon}, V., {et~al.} 2000, \apj, 531, 1003, \dodoi{10.1086/308491}

\bibitem[{{Rivinius} {et~al.}(2013){Rivinius}, {Carciofi}, \& {Martayan}}]{rivinius}
{Rivinius}, T., {Carciofi}, A.~C., \& {Martayan}, C. 2013, \aapr, 21, 69, \dodoi{10.1007/s00159-013-0069-0}

\bibitem[{{Rivinius} \& {Klement}(2024)}]{klement24}
{Rivinius}, T., \& {Klement}, R. 2024, arXiv e-prints, arXiv:2411.06882, \dodoi{10.48550/arXiv.2411.06882}

\bibitem[{{Rivinius} {et~al.}(2006){Rivinius}, {{\v{S}}tefl}, \& {Baade}}]{rivi06}
{Rivinius}, T., {{\v{S}}tefl}, S., \& {Baade}, D. 2006, \aap, 459, 137, \dodoi{10.1051/0004-6361:20053008}

\bibitem[{{Rizzuto} {et~al.}(2013){Rizzuto}, {Ireland}, {Robertson}, {Kok}, {Tuthill}, {Warrington}, {Haubois}, {Tango}, {Norris}, {ten Brummelaar}, {Kraus}, {Jacob}, \& {Laliberte-Houdeville}}]{2013MNRAS.436.1694R}
{Rizzuto}, A.~C., {Ireland}, M.~J., {Robertson}, J.~G., {et~al.} 2013, \mnras, 436, 1694, \dodoi{10.1093/mnras/stt1690}

\bibitem[{{Ruban} {et~al.}(2006){Ruban}, {Alekseeva}, {Arkharov}, {Hagen-Thorn}, {Galkin}, {Nikanorova}, {Novikov}, {Pakhomov}, \& {Puzakova}}]{ruban}
{Ruban}, E.~V., {Alekseeva}, G.~A., {Arkharov}, A.~A., {et~al.} 2006, Astronomy Letters, 32, 604, \dodoi{10.1134/S1063773706090052}

\bibitem[{{Sana} {et~al.}(2009){Sana}, {Gosset}, \& {Evans}}]{sana09}
{Sana}, H., {Gosset}, E., \& {Evans}, C.~J. 2009, \mnras, 400, 1479, \dodoi{10.1111/j.1365-2966.2009.15545.x}

\bibitem[{{Scardia} {et~al.}(2006){Scardia}, {Prieur}, {Pansecchi}, {Argyle}, {Sala}, {Ghigo}, {Koechlin}, \& {Aristidi}}]{scardia}
{Scardia}, M., {Prieur}, J.~L., {Pansecchi}, L., {et~al.} 2006, \mnras, 367, 1170, \dodoi{10.1111/j.1365-2966.2006.10035.x}

\bibitem[{{Schmidt}(1968)}]{vmax}
{Schmidt}, M. 1968, \apj, 151, 393, \dodoi{10.1086/149446}

\bibitem[{{Scott} {et~al.}(2021){Scott}, {Howell}, {Gnilka}, {Stephens}, {Salinas}, {Matson}, {Furlan}, {Horch}, {Everett}, {Ciardi}, {Mills}, \& {Quigley}}]{zorro}
{Scott}, N.~J., {Howell}, S.~B., {Gnilka}, C.~L., {et~al.} 2021, Frontiers in Astronomy and Space Sciences, 8, 138, \dodoi{10.3389/fspas.2021.716560}

\bibitem[{{Skiff}(2010)}]{skiff}
{Skiff}, B.~A. 2010, {VizieR Online Data Catalog: Catalogue of Stellar Spectral Classifications (Skiff, 2009-2012)}, VizieR On-line Data Catalog: B/mk. Originally published in: 2009yCat....1.2023S

\bibitem[{{Souza} {et~al.}(2020){Souza}, {Guerrero}, \& {Borges Fernandes}}]{2020AJ....159..132S}
{Souza}, T.~B., {Guerrero}, C.~A., \& {Borges Fernandes}, M. 2020, \aj, 159, 132, \dodoi{10.3847/1538-3881/ab6dcd}

\bibitem[{{Suffak} {et~al.}(2025){Suffak}, {Jones}, \& {Carciofi}}]{belzldv}
{Suffak}, M.~W., {Jones}, C.~E., \& {Carciofi}, A.~C. 2025, \mnras, 536, 2234, \dodoi{10.1093/mnras/stae2709}

\bibitem[{{Tokovinin}(2023)}]{toko23}
{Tokovinin}, A. 2023, \aj, 165, 180, \dodoi{10.3847/1538-3881/acc464}

\bibitem[{{Tokovinin} {et~al.}(2010){Tokovinin}, {Mason}, \& {Hartkopf}}]{toko10}
{Tokovinin}, A., {Mason}, B.~D., \& {Hartkopf}, W.~I. 2010, \aj, 139, 743, \dodoi{10.1088/0004-6256/139/2/743}

\bibitem[{{Tokovinin} {et~al.}(2021){Tokovinin}, {Mason}, {Mendez}, {Costa}, {Mann}, \& {Henry}}]{toko21}
{Tokovinin}, A., {Mason}, B.~D., {Mendez}, R.~A., {et~al.} 2021, \aj, 162, 41, \dodoi{10.3847/1538-3881/ac00bd}

\bibitem[{{Vioque} {et~al.}(2020){Vioque}, {Oudmaijer}, {Schreiner}, {Mendigut{\'\i}a}, {Baines}, {Mowlavi}, \& {P{\'e}rez-Mart{\'\i}nez}}]{vioque}
{Vioque}, M., {Oudmaijer}, R.~D., {Schreiner}, M., {et~al.} 2020, \aap, 638, A21, \dodoi{10.1051/0004-6361/202037731}

\bibitem[{{Wang} {et~al.}(2017){Wang}, {Gies}, \& {Peters}}]{wang17}
{Wang}, L., {Gies}, D.~R., \& {Peters}, G.~J. 2017, \apj, 843, 60, \dodoi{10.3847/1538-4357/aa740a}

\bibitem[{{Wang} {et~al.}(2018){Wang}, {Gies}, \& {Peters}}]{wang18}
---. 2018, \apj, 853, 156, \dodoi{10.3847/1538-4357/aaa4b8}

\bibitem[{{Wang} {et~al.}(2023){Wang}, {Gies}, {Peters}, \& {Han}}]{wang23}
{Wang}, L., {Gies}, D.~R., {Peters}, G.~J., \& {Han}, Z. 2023, \aj, 165, 203, \dodoi{10.3847/1538-3881/acc6ca}

\bibitem[{{Wood} {et~al.}(2021){Wood}, {Mann}, \& {Kraus}}]{molusc}
{Wood}, M.~L., {Mann}, A.~W., \& {Kraus}, A.~L. 2021, \aj, 162, 128, \dodoi{10.3847/1538-3881/ac0ae9}

\bibitem[{{Zheng} {et~al.}(2023){Zheng}, {Cao}, {Deng}, {Mei}, {Tan}, \& {Wang}}]{lamost}
{Zheng}, Z., {Cao}, Z., {Deng}, H., {et~al.} 2023, \apjs, 266, 18, \dodoi{10.3847/1538-4365/acc94e}

\end{thebibliography}
\bibliographystyle{aasjournal}



\end{document}